\documentclass[apj]{emulateapj}

\usepackage{natbib}
\usepackage{color}
\usepackage{mediabb}
\usepackage{amsmath,amssymb}

\newcommand{\myemail}{hideki.umehata@ouj.ac.jp}

\slugcomment{draft November 16 2016}

\shorttitle{ADF22: Catalog and Counts}
\shortauthors{Umehata et al.}

\begin{document}


\title{ALMA Deep Field in SSA22: Source Catalog and Number Counts}


\author{Hideki Umehata\altaffilmark{1,2,3}, Yoichi Tamura\altaffilmark{2}, Kotaro Kohno\altaffilmark{2,4}, R.J. Ivison\altaffilmark{3,5}, Ian Smail\altaffilmark{6}, Bunyo Hatsukade\altaffilmark{7}, Kouichiro Nakanishi\altaffilmark{7,8}, Yuta Kato\altaffilmark{7,9}, Soh Ikarashi\altaffilmark{10}, Yuichi Matsuda\altaffilmark{7,8}, Seiji Fujimoto\altaffilmark{11}, Daisuke Iono\altaffilmark{7,8}, Minju Lee\altaffilmark{7,9}, Charles C. Steidel\altaffilmark{12}, Tomoki Saito\altaffilmark{7}, D.M. Alexander\altaffilmark{6}, Min S. Yun\altaffilmark{13}, Mariko Kubo\altaffilmark{7,11}}
\affil{
$^{1}$ The Open University of Japan, 2-11 Wakaba, Mihama-ku, Chiba 261-8586, Japan;
\textcolor{blue}{{\myemail}}
\\
$^2$ Institute of Astronomy, School of Science, The University of Tokyo, 2-21-1 Osawa, Mitaka, Tokyo 181-0015, Japan
\\
$^3$ European Southern Observatory, Karl-Schwarzschild-Str. 2, D-85748 Garching, Germany
\\
$^4$ Research Center for the Early Universe, The University of Tokyo, 7-3-1 Hongo, Bunkyo, Tokyo 113-0033
\\
$^5$ Institute for Astronomy, University of Edinburgh, Royal Observatory, Blackford Hill, Edinburgh EH9 3HJ, UK
\\
$^6$ Centre for Extragalactic Astronomy, Department of Physics, Durham University, South Road, Durham, DH1 3LE, UK
\\
$^7$ National Astronomical Observatory of Japan, 2-21-1 Osawa, Mitaka, Tokyo 181-8588, Japan
\\
$^8$ Department of Astronomy, School of Science, SOKENDAI (The Graduate University for Advanced Studies), Osawa, Mitaka,
Tokyo 181-8588, Japan
\\
$^{9}$Department of Astronomy, Graduate school of Science, The University of Tokyo, 7-3-1 Hongo, Bunkyo-ku, Tokyo 133-0033, Japan
\\
$^10$ Kapteyn Astronomical Institute, University of Groningen, P.O. Box 800, 9700AV Groningen, The Netherlands
\\
$^{11}$ Institute for Cosmic Ray Research, The University of Tokyo, Kashiwa, Chiba 277-8582, Japan
\\
$^{12}$ California Institute of Technology, MS 249-17, Pasadena, CA 91125, USA
\\
$^{13}$ Department of Astronomy, University of Massachusetts, Amherst, MA 01003, USA
}

%




\begin{abstract}
We present results from a deep 2$^\prime \times 3^\prime$ (comoving scale of 3.7~Mpc$\times$ 5.5~Mpc at $z=3$) survey  at 1.1~mm taken with the Atacama Large Millimeter/submillimeter Array (ALMA) in the SSA22 field.
We observe the core region of a $z=3.09$ protocluster, achieving a typical rms sensitivity of 60~$\mu$Jy beam$^{-1}$ at a spatial resolution of $0^{\prime\prime}.7$. 
We detect 18 robust ALMA sources at a signal-to-noise ratio (SNR) $>5$.
Comparison between the ALMA map and a 1.1~mm map taken with the AzTEC camera on the Atacama Submillimeter Telescope Experiment (ASTE) indicates that three submillimeter sources discovered by the AzTEC/ASTE survey are resolved into eight individual submillimeter galaxies (SMGs) by ALMA.
At least ten of our 18 ALMA SMGs have spectroscopic redshifts of $z\simeq3.09$, placing them in the protocluster.
This shows that a number of dusty starburst galaxies are forming simultaneously in the core of the protocluster.
The nine brightest ALMA SMGs with SNR $>10$ have a median intrinsic angular size of $0^{\prime\prime}.32^{+0.13}_{-0.06}$ ($2.4^{+1.0}_{-0.4}$ physical kpc at $z=3.09$), which is consistent with previous size measurements of SMGs in other fields.
As expected the source counts show a possible excess compared to the counts in the general fields at $S_{\rm 1.1 mm} \ge1.0$~mJy due to the protocluster.
Our contiguous mm mapping highlights the importance of large-scale structures on the formation of dusty starburst galaxies.
\end{abstract}





\section{Introduction}

The history of galaxy formation and evolution appears to be linked to the growth of cosmic large-scale structure.
In the present-day Universe, clusters of galaxies represent some of the densest environments.
Massive and passive elliptical galaxies preferentially reside in cluster cores (e.g., \citealt{1980ApJ...236..351D}; \citealt{1984ApJ...281...95P}).
Recent works have suggested that brightest cluster galaxies at the center of clusters at $z=1-2$ are relatively evolved and 
hence they experienced rapid growth at earlier epochs (e.g., \citealt{2009Natur.458..603C}; \citealt{2011A&A...526A.133G}).
Galaxy cluster archaeology and cosmological simulations suggest the rapid growth of massive ellipticals in high density regions at $z\gtrsim2-3$ (e.g., \citealt{2005ApJ...632..137N}; \citealt{2006MNRAS.366..499D}).
Therefore uncovering intense star-forming activity in protoclusters at such high redshifts is of great importance for understanding galaxy and cluster formation in the era when they actively assembled their stars.
Such actively star-forming galaxies are expected to be enshrouded by dust, which renders them difficult to observe at rest-frame UV to optical wavelengths. 
Their spectral energy distributions (SEDs) thus should be dominated by far-infrared (FIR) emission, and therefore observing at FIR to submm/mm wavelengths may be essential to uncover these dusty galaxies (so called submillimeter galaxies or SMGs; for reviews, see \citealt{2002PhR...369..111B}; \citealt{2014PhR...541...45C}).

Extensive efforts have been made to search for and identify such obscured star-forming galaxies in protoclusters at $z>2$, using submm/mm bolometer cameras onboard single-dish telescopes, such as AzTEC (e.g., \citealt{2009Natur.459...61T}; \citealt{2011Natur.470..233C}; \citealt{2014MNRAS.440.3462U}), SCUBA (e.g., \citealt{2004ApJ...611..725B}; \citealt{2009ApJ...694.1517D}), SCUBA2 (e.g., \citealt{2015ApJ...808L..33C}), and LABOCA (e.g., \citealt{2014A&A...570A..55D}; \citealt{2016MNRAS.tmp..952C}) as well as FIR to submm/mm satellites, including  {\it Herschel} (e.g., \citealt{2016MNRAS.460.3861K}) and {\it Planck} (e.g., \citealt{2015A&A...582A..30P}). 
While these single-dish telescopes are beneficial to cover wide area and find bright SMGs, their relatively poor angular resolution (FWHM $\gtrsim15^{\prime\prime}-30^{\prime\prime}$) and sensitivity limits due to source confusion have prevented us from obtaining accurate identifications for sources and/or revealing 
less extreme dusty galaxies.
The advent of the Atacama Large Millimeter/submillimeter Array (ALMA) allows us to break through these limitations.
Contiguous ALMA mosaic imaging is able to open a window for submm/mm deep surveys with sub-arcsec resolution (e.g., \citealt{2015ApJ...811L...3T}; \citealt{2016PASJ...68...36H}; \citealt{2016arXiv160600227D}; \citealt{2016arXiv160706769A}; \citealt{2016arXiv160706768W}).
\footnote{
In order to distinguish the submm/mm sources discovered by single-dish telescopes and the galaxies individually observed by interferometers clearly, we utilize the word, SMGs, to indicate the latter -- the galaxies individually identified at submm/mm wavelengths -- in this paper.
We consider all sources discovered by our ALMA survey as SMGs.
As we will show, the individual ALMA SMGs have infrared luminosities comparable with ULIRGs for most cases.
The sources discovered by single-dish surveys are simply called ``sources'' (e.g., AzTEC sources).
}

We utilized ALMA to conduct a deep imaging survey toward a well-studied protocluster at $z=3.09$ in the SSA22 field.
The protocluster was originally discovered as a redshift spike of Lyman-break galaxies (LBGs) and Ly$\alpha$ emitters (LAEs) by \citet{1998ApJ...492..428S}, (\citeyear{2000ApJ...532..170S}) and was proposed as an ancestor of present-day clusters such as Coma.
A remarkable LAE density peak ($\sim$6 times the average surface density) spreading over tens of comoving Mpc has been found among a huge filamentary structure ($>100$ comoving Mpc) traced by LAEs (\citealt{2000ApJ...532..170S}; \citealt{2004AJ....128.2073H}; \citealt{2005ApJ...634L.125M}; \citealt{2012AJ....143...79Y}).
Distant red galaxies (DRGs) and $K$-band selected galaxies are found to be more abundant in the core of the protocluster than the field, which supports the elevated formation of massive galaxies there (\citealt{2012ApJ...750..116U}; \citealt{2013ApJ...778..170K}; \citeyear{2015ApJ...799...38K}).
In the SSA22 field, a number of submm/mm surveys have been conducted (e.g., \citealt{2005MNRAS.363.1398G}, \citeyear{2014ApJ...793...22G}; \citealt{2006MNRAS.370.1057S}; \citealt{2009Natur.459...61T}; \citealt{2014MNRAS.440.3462U}).
\citet{2009Natur.459...61T} discovered a statistical correlation between 15 bright 1.1~mm sources detected by AzTEC and $z\sim3.09$ LAEs and suggested that SMGs preferentially formed within the cosmic structure.
\citet{2014MNRAS.440.3462U} performed counterpart identification of AzTEC sources using radio and near-to-mid infrared data and derived optical to near-infrared photometric redshifts ($z_{\rm phot}$), which supported the trend found in \citet{2009Natur.459...61T} and
led to the conclusion that a significant fraction of the AzTEC sources are concentrated in the center of the protocluster.

On the basis of our ALMA survey, in \citet{2015ApJ...815L...8U} we reported the discovery of a concentration of dusty starbursts and X-ray AGNs at a center of the protocluster, and suggested that the large-scale environment plays a key role in the formation of these rare and active populations (see also \citealt{2016MNRAS.461.2944A}). Here we present the full catalog of the ALMA SMGs and the comprehensive results in terms of mm properties unveiled by ALMA.
Our survey design and observations are described in \S 2. We present the procedures for source extraction and catalog selection in \S3.
We compare the ALMA survey with the previous AzTEC survey and derive properties of the sources including source counts in \S4.
We conclude with a summary in \S 5.
We will report the result of a line search in our survey in Hayatsu et al. (in preparation).
Throughout the paper, we adopt a cosmology with 
$\Omega_{\rm m}=0.3, \Omega_\Lambda=0.7$, and H$_0$=70 km s$^{-1}$ Mpc$^{-1}$.

\section{The ALMA Data}

\subsection{Field Selection}

\begin{figure*}
\epsscale{0.95}
\plotone{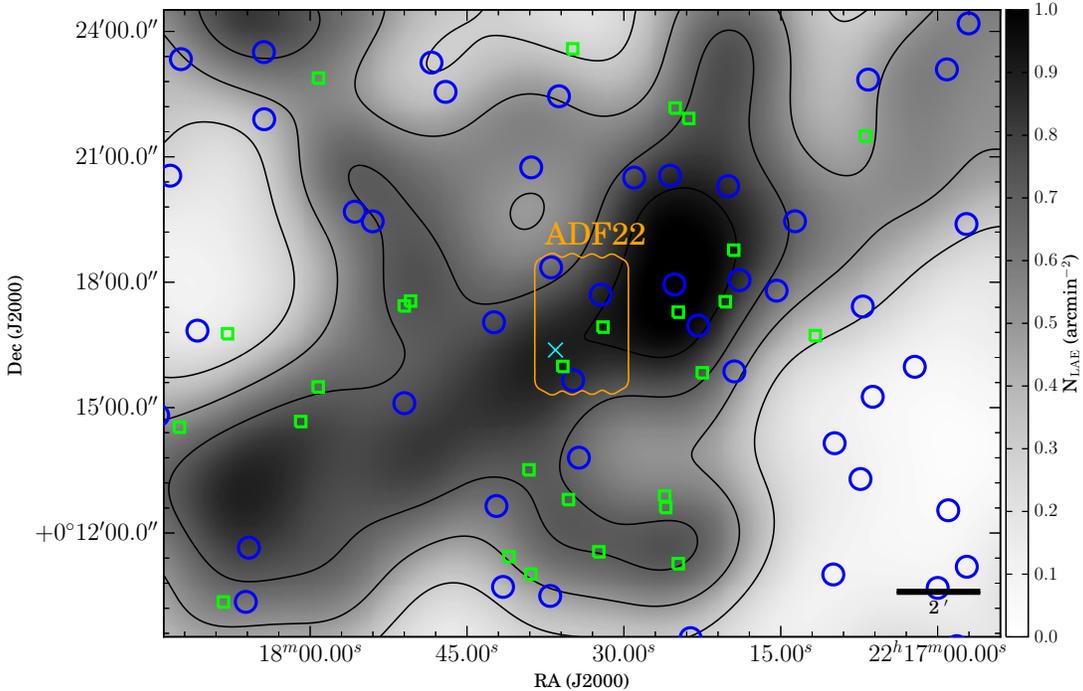}
\caption{
The location of the observed area (named as ADF22) on the $z=3.09$ LAE candidate density map (\citealt{2004AJ....128.2073H}).
The orange line shows the area
we mapped with ALMA to a limit corresponding to 50\% sensitivity of the final mosaic map (same as Fig.~\ref{area}).
The gray scales show the surface number density of LAE candidates at $z\sim3.09$ and the contours represent the smoothed LAE density of 0.3, 0.5, 0.7, and 0.9 in units of arcmin$^{-2}$, which outlines the $z=3.1$ protocluster (\citealt{2004AJ....128.2073H}).
The blue circles show the 1.1 mm sources discovered by the AzTEC/ASTE survey ($\ge 3.5 \sigma$; \citealt{2014MNRAS.440.3462U}). Their diameters are 30 arcsec, which is equivalent to the angular resolution of the AzTEC map.
The green squares show the $z\sim3.09$ LAB positions (\citealt{2004AJ....128..569M}) and the cyan cross indicates a $z=3.09$ QSO (\citealt{1998ApJ...492..428S}).
ADF22 is located at close to the LAE density peak and contains three submm sources, two bright LABs, and one QSO.
}
\label{f1}
\end{figure*}

We observed a $2^{\prime} \times 3^{\prime}$ rectangular field and peripheral regions centered at $\alpha = 22^{\rm h}17^{\rm m}34^{s}.0$, $\delta = + 00^\circ 17^\prime 00^{\prime\prime}$ (J2000) at 1.1~mm, using ALMA Band 6 in cycle 2 (Program ID. 2013.1.00162.S, PI. H. Umehata).
Hereafter we name this field ``ADF22'', which is an acronym of ``ALMA deep field in the SSA22 field''.
ADF22 is located at the very center of the cosmic large-scale structure (or ``Cosmic Web'') at $z\sim3$.
As illustrated in Fig.~\ref{f1}, this area is close to the projected density peak of $z=3.09$ LAE candidates selected by Subaru/Suprime-cam observations with a narrow-band filter (NB497) (\citealt{2004AJ....128.2073H}).
Previous works have unveiled not only the projected density distribution but the density structure in three-dimensions (\citealt{2005ApJ...634L.125M}).
When we only focus on spectroscopically confirmed $z=3.09$ LAEs and draw a three-dimensional map, ADF22 coincides with the junction of a comoving 50 Mpc-scale filamentary structure (\citealt{2015ApJ...815L...8U}).
This means that ADF22 covers a node of the cosmic large-scale structure and provides a unique opportunity to examine galaxy formation in such a dense field at this early epoch.

Another important feature is that ADF22 contains a significant numbers of the most active populations of galaxies which are suggested to preferentially reside in dense environments, including SMGs, QSOs, and Lyman alpha blobs (LABs).
There are three 1.1~mm sources discovered by the AzTEC/ASTE survey (\citealt{2009Natur.459...61T}; \citealt{2014MNRAS.440.3462U}). 
Prior to the present work, one of the three sources, SSA22-AzTEC77 (or SMM J221735+001537), had been proposed as a secure member of the $z=3.09$ structure (\citealt{2005MNRAS.359.1165G}; \citealt{2005ApJ...622..772C}; \citealt{2013MNRAS.429.3047B}) and the other two sources had been also suggested to be at $z\sim3.09$ (\citealt{2010ApJ...724.1270T}; \citealt{2012ApJ...750..116U}; \citealt{2014MNRAS.440.3462U}).
One QSO at $z=3.09$ discovered by \citet{1998ApJ...492..428S} and two $z\sim3.09$ LABs listed in \citet{2004AJ....128..569M} are also located within ADF22 (see also Fig.~\ref{f1}).
ADF22 is thus an excellent region for probing dusty star-formation activity in a wide variety of galaxies at the core of the protocluster.

\subsection{Observations}

We chose the central observing frequency of 263~GHz (1.14~mm), which is similar to that of our previous AzTEC/ASTE survey, 270 GHz, and 
allows direct comparison with the flux densities of sources from our AzTEC/ASTE and ALMA surveys.
Since the size of primary beam of band~6 is larger than that of band~7, the frequency set-up is also beneficial to reduce the number of pointings compared to higher frequency.
In order to cover a $2^{\prime}\times3^{\prime}$ rectangular field contiguously, we utilized a 103-field mosaic.
The spacing between adjacent pointings was 0.72 times the primary beam (FWHM 23$^{\prime\prime}$ at 263~GHz), which was a compromise between the homogeneity of sensitivity and the observing time required to cover as wide an area as possible.

Observations were divided into two campaigns (2014 and 2015 runs), as summarized in Table 1.
The first run was conducted on 2014 June 6--10 with 33--36 available 12 m antennas in the C34-4 configuration (longest baseline 650~m) and very good 1.1~mm weather conditions (precipitable water vapor (PWV) of 0.30--1.27~mm).
The second run was done on 2015 April 4, 5, and 13 with 32--40 12m antennas in C32-2 (longest baseline 349~m) and good conditions (PWV of 0.9--1.9~mm).
The exposure time per pointing was 30~sec for each scheduling block (SB).
These observations resulted in initial on-source time of 2--4.5~min per pointing (depending on the individual pointing) and a total on-source time of 386 min.
We utilized the Time Division Modes (TDM) correlator, with $4\times128$ dual polarization channels over the full 8~GHz bandwidth, giving an effective bandwidth of 7.5~GHz after flagging edge channels.
The correlator was set up to target two spectral windows of 1.875 GHz bandwidth each at 15.6~MHz ($\sim$ 20 km s$^{-1}$) channel spacing in each sideband.
The central frequencies of the four spectral windows are 254.0, 256.0, 270.0, and 272.0 GHz.
We note that this frequency range enables us to search for the $^{12}$CO(9-8) line ($\nu_{\rm rest}=$1036.912 GHz) at $z\sim3.09$ 
(we actually found one CO(9-8) emitter as we describe below. Details will be presented in Hayatsu et al., in preparation.).

The quasar J2148+0657  with a flux 1.2 Jy was observed regularly for amplitude and phase calibration.
The absolute flux scale was set using J2148+0657 (for the 2014 run) and Uranus and Neptune (for the 2015 run).
We estimate that the absolute flux accuracy is within 20\%.
This uncertainty in the absolute flux calibration is not included in the following analyses and discussions.

\begin{center}
\begin{deluxetable*}{ccccccc}
\tabletypesize{\scriptsize}
\tablecaption{Summary of ADF22 Observations}
\tablewidth{0pt}
\tablehead{
\colhead{SB$^{a}$} &\colhead{Date} & \colhead{Antennas$^{b}$} & \colhead{Baseline$^{c}$} & \colhead{Fields$^{d}$} & \colhead{Synthesized beam$^{e}$} & \colhead{Flux calibrator}}   
\startdata
SB1 & 2014 Jun 06 & 36 & 20 m -- 650 m & 1--80 & $0.59^{\prime\prime}\times0.46^{\prime\prime}$ ($-37.5$ deg) &   J2148+069   \\   
SB2 & 2014 Jun 07 & 34 & 20 m -- 646 m &1--40 & $0.50^{\prime\prime}\times0.45^{\prime\prime}$  ($+18.6$ deg) & J2148+069   \\   
SB3 & 2014 Jun 07 & 34 & 20 m -- 646 m &1--80 & $0.56^{\prime\prime}\times0.47^{\prime\prime}$  ($-59.6$ deg) &   J2148+069   \\   
SB4 & 2014 Jun 09 & 34 & 20 m -- 646 m &1--80 & $0.61^{\prime\prime}\times0.46^{\prime\prime}$  ($+63.7$ deg) &   J2148+069  \\   
SB5 & 2014 Jun 10 & 34 & 20 m -- 646 m &1--80 & $0.57^{\prime\prime}\times0.48^{\prime\prime}$  ($-64.9$ deg) &  J2148+069 \\   
SB6 & 2015 Apr 04 & 33 & 15 m -- 328 m &1--103 & $1.18^{\prime\prime}\times0.87^{\prime\prime}$  ($-88.6$ deg) & Neptune  \\   
SB7 & 2015 Apr 04 & 33 & 15 m -- 328 m &1--103 & $1.39^{\prime\prime}\times0.81^{\prime\prime}$ ($-70.1$ deg) &  Neptune  \\   
SB8 & 2015 Apr 05 & 35 & 15 m -- 328 m & 1--103 & $1.30^{\prime\prime}\times0.90^{\prime\prime}$  ($-72.3$ deg) &  Uranus  \\   
SB9 & 2015 Apr 13 & 40 & 15 m -- 349 m &1--103 & $1.67^{\prime\prime}\times1.11^{\prime\prime}$  ($+73.9$ deg) &  Neptune   
\enddata
\tablecomments{
$^{a}$ Scheduling block; 
$^{b}$ Number of utilized 12~m antennas; 
$^{c}$ Minimum and maximum baseline;
$^{d}$ Field ID of each pointing (see Fig.~\ref{f2});
$^{e}$ Synthesized beam size of the map in the case of natural weighting
}
\end{deluxetable*}
\end{center}

\subsection{Data Reduction and Imaging}

\begin{figure*}
\epsscale{0.95}
\plotone{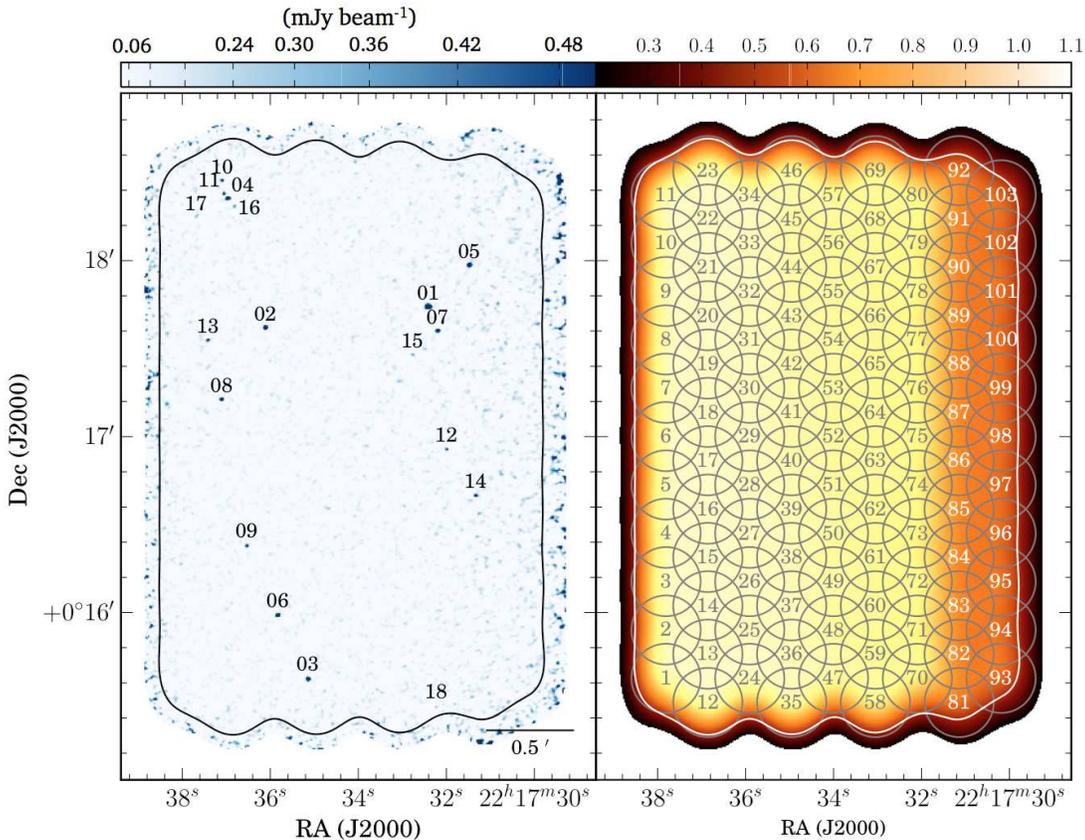}
\caption{
(left) Our ALMA 1.1~mm ``FULL/LORES'' map of ADF22, corrected for primary beam response. 
The solid black contour shows the border at which the flux attenuation is 50\%.
Source extraction was conducted within the 50\% coverage area.
We also mark the 5$\sigma$ source positions with identifications (IDs). 
(right) The ``flux'' map of ADF22 created by {\sc casa}, which shows the response function of flux attenuation within the area.
White contours shows the 50\% attenuation as in the left figure. 
Since the ADF22 consists of observations from 103 discrete fields, we denote them as shown in the figure (fields 1--103).
The ``DEEP/HIRES'' map is also created using only 80 pointings (fields 1--80; see \S2.3.).
Each circle show an individual field of view, corresponding to a single pointing ($d=23^{\prime\prime}$).
For both maps, the outer contour shows the 20\% coverage area.
}
\label{f2}
\end{figure*}

\begin{figure}
\epsscale{1.15}
\plotone{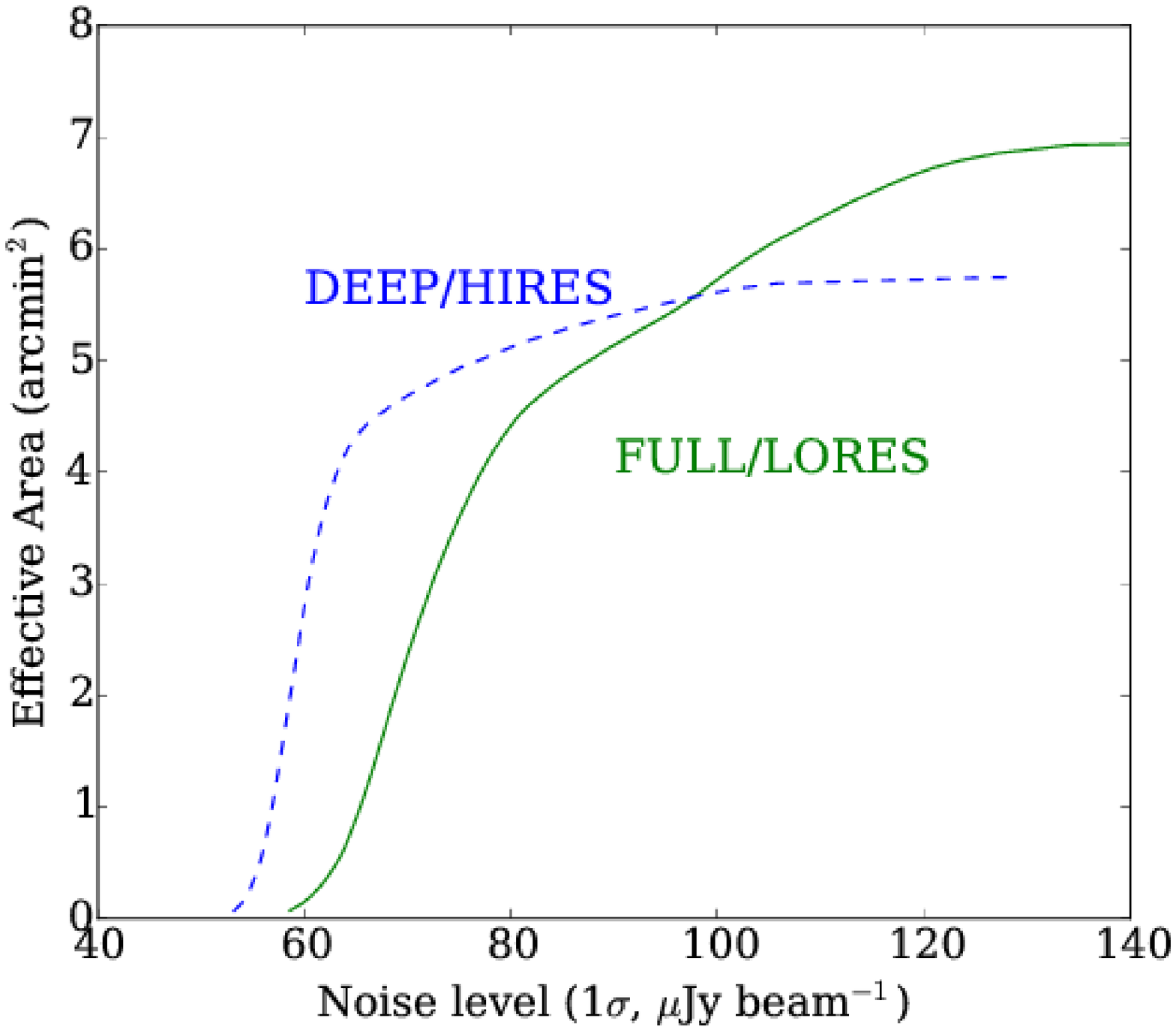}
\caption{
Effective survey area of the FULL/LORES map and the DEEP/HIRES map as a function of 1$\sigma$ sensitivity.
Our survey covers 7.0 arcmin$^2$ and 5.8 arcmin$^2$ in total, respectively. 
}
\label{area}
\end{figure}

Data reduction was performed using the Common Astronomy Software Application ({\sc casa}; \citealt{2007ASPC..376..127M})\footnote{http://casa.nrao.edu}.
The calibration and imaging was performed using {\sc casa} version 4.2.2 of the ALMA pipeline and {\sc casa} version 4.5.0, respectively.
Our analysis is complicated by the fact that our observations were carried out with multiple array configurations.
Our observations are composed of nine SBs as summarized in Table 1.
Since each SB has a different antenna setup, the resulting synthesized beam sizes differ among SBs.
There are broadly two array configurations; the maximum baseline lengths of SB6--SB9 (taken in 2015) are systematically shorter than SB1--SB5 (taken in 2014).
Because of the difference in the observed area in 2014 and 2015, the western edge fields (fields 81--103 in the right panel of Fig.~\ref{f2}) lack longer baseline data and have lower sensitivity (Table 1).

Considering this situation, we first created a 1$^{\prime\prime}$-tapered map for the entire field (i.e., fields 1--103), which enables us to achieve nearly uniform angular resolution across the map.
This larger synthesized beam is more sensitive to extended sources but suffers from an increased noise level. 
As we show in the maps of individual sources (Fig.~\ref{stamp}) and measured sizes of bright sources (Table 3), the resulting angular resolution, $\approx1^{\prime\prime}$, seems to be sufficient to detect the majority of sources in this field.
We also discuss the possibility that we are missing very extended components in \S 4.1.1.

The $u v$-data for individual pointings were first combined into a single $u v$--data set.
We then Fourier-transformed the combined data to create a single ``dirty'' map with natural weighting, by setting the imager mode to ``mosaic'' and adopting the taper parameter, outertaper = 120 k$\lambda$. 
After we measured the rms noise level across the whole dirty map, we repeated the clean process down to 1.5$\sigma$, putting a tight clean box ($1^{\prime\prime}\times1^{\prime\prime}$ in size) around each 5$\sigma$ source (in a manner similar to that reported by \citealt{2013ApJ...768...91H}).
The resulting synthesized beam is $1.^{\prime\prime}16\times1.^{\prime\prime}02$ in size (P.A. $= -80$ deg).
We denote this map as the FULL/LORES map.

We created another complementary map.
Since the five SBs, SB1 -- SB5, have the longer baseline data with baselines up to 600 k$\lambda$, 
we can achieve a better sensitivity making the best of the longer baselines.
For this purpose, we created a second map for 80 pointing field (fields1--80).
We in general adopted the same procedure described above, but we did not apply any tapering to the map.
The second map has a synthesized beam of $0.^{\prime\prime}70\times0.^{\prime\prime}59$ (P.A. $= -80$ deg).
Hereafter we call it the DEEP/HIRES map.
In the following analyses, we utilize primarily the FULL/LORES map.
The DEEP/HIRES map is used for detecting compact faint sources (\S 3) and resolving bright sources  (\S 4.2).

The final FULL/LORES map corrected for the primary beam response is shown in Fig.~\ref{f2}.
In the following sections, we consider the effective area; the area in the map within which the primary beam coverage is greater than 50\%.
This results in 7.0 arcmin$^2$ area in the case of the FULL/LORES map (5.8 arcmin$^{2}$ for the DEEP/HIRES map).
A sensitivity map was constructed utilizing an {\sc aips} (\citealt{2003ASSL..285..109G}) task, {\sc rmsd}, calculating the rms for each $0.^{\prime\prime}1\times0.^{\prime\prime}1$ pixel using the surrounding $100\times100$ pixels (or $10.^{\prime\prime}0\times10.^{\prime\prime}$0) on an image prior to correcting for primary beam response.
The correction results in a range of 1$\sigma$ depth of 52 -- 170 $\mu$Jy beam$^{-1}$ with a median value of 75 $\mu$Jy beam$^{-1}$ (Fig.~\ref{area}) for the FULL/LORES map.
In the case of the DEEP/HIRES map, we obtained a range of 1$\sigma$ depth from 50 -- 129 $\mu$Jy beam$^{-1}$ with a median value of 60 $\mu$Jy beam$^{-1}$ (Fig.~\ref{area}).

\section{The Catalog}

\subsection{Source Extraction and Characterization}

\begin{figure}
\epsscale{1.15}
\plotone{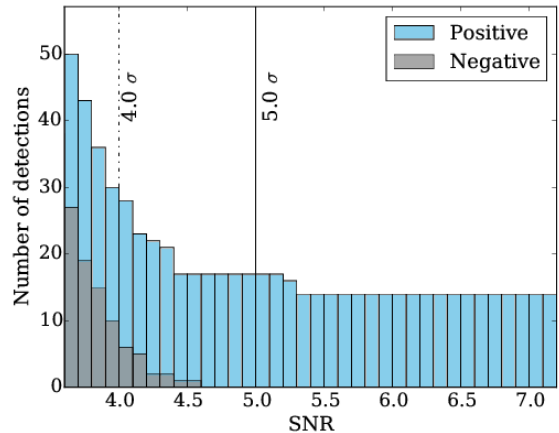}
\caption{
Cumulative number of positive and negative peaks as a function of signal-to-noise ratio (SNR) for the FULL/LORES map.
The solid vertical line shows the detection limit ($5\sigma$).
There are no negative peaks above 5$\sigma$, which suggests our adopted detection limit is conservative.
The dash-dot line shows the threshold for tentative detection.
}
\label{snr}
\end{figure}

Source extraction was performed using the image map and sensitivity map that were not corrected for the primary beam attenuation. 
First we utilized the FULL/LORES map.
A source-finding algorithm, {\sc aegean} v1.9.5-56 (\citealt{2012MNRAS.422.1812H}), was used to extract positive sources above 3.5$\sigma$.
In parallel, we also counted {\it negative} sources with source-like profiles above 3.5$\sigma$ during the procedure to estimate the fraction of spurious sources among the detected sources.
The cumulative number counts of positive and negative sources as a function of signal-to-noise ratio (SNR) is shown in Fig.~\ref{snr}.
We found 17 positive sources with $>5\sigma$ and 28 sources with $>4\sigma$.
There were no negative sources above 5$\sigma$, while we found seven negative ones with 4--4.6$\sigma$.
The results show that the detection limit of 5$\sigma$ is secure and conservative.
Gaussian statistics also support the validity of the threshold. Since the map contains $\approx$32000 beams, we would expect less than one spurious peaks above 5$\sigma$.
Therefore we adopt these 17 sources as secure detections (ADF22.1--ADF22.17; see Table 2).

We also performed the same procedures on the DEEP/HIRES map, independently.
As a result, 14 sources were detected at $>5\sigma$, one of which was not detected in the shallower FULL/LORES map.
This source (ADF22.18; see Table 2) was added to the catalog of secure detection 
because there were no negative sources with $>5\sigma$ in the DEEP/HIRES map. 
Therefore in total 18 sources were detected at SNR$>5$.

In the following analyses and discussion, we mostly focus on the 18 secure sources not affected by spurious features.
We show a supplementary catalog of 14 sources detected tentatively in the FULL/LORES and DEEP/HIRES maps in Appendix~A.

The flux densities of the detected sources were measured with the {\sc imfit} task of {\sc casa}, using the map after correction for the primary beam attenuation.
We adopt the integrated flux density as the flux density of a source unless it is lower than the peak flux density (e.g., \citealt{2015ApJ...799...81S}).
The measurements of fluxes on the two maps are in good agreement with each other.
The median ratio of the flux density between the FULL/LORES map and the DEEP/HIRES map is $S^{\rm FULL}/S^{\rm DEEP}=0.96^{+0.00}_{-0.02}$ for the 13 sources detected above $5\sigma$ in both maps.

\subsection{Completeness and Flux boosting}

\begin{figure}
\epsscale{1.15}
\plotone{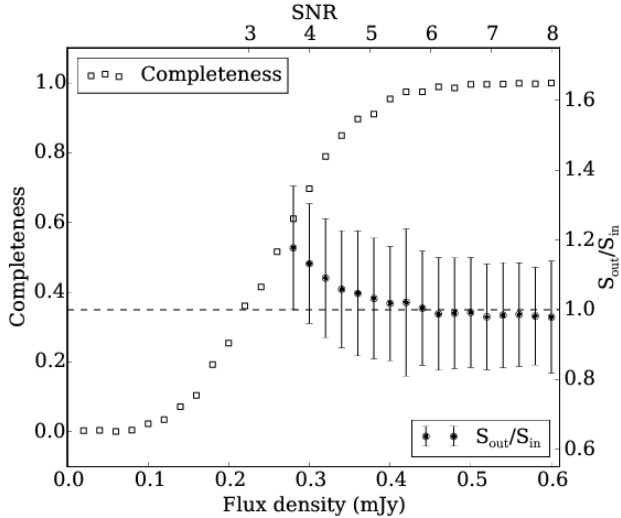}
\caption{
Measured completeness and the ratio between input flux densities of embedded artificial sources and output flux densities for the FULL/LORES map.
The lower horizontal axes shows input flux density, corrected for primary beam attenuation.
The upper axis shows the corresponding, approximate SNR, calculated assuming the median sensitivity.
The $5\sigma$ detection limit results in $\gtrsim$90\% completeness.
We do not consider the flux boosting effect, which is negligible for $5\sigma$ sources.
}
\label{f4}
\end{figure}

We performed a suite of simulations to evaluate the completeness and boosting effect on our flux measurement.
We briefly describe the method below (for more detail, see \citealt{2016PASJ...68...36H}).
Model sources created by scaling the synthesized beam are injected into the image map corrected for primary beam attenuation.
The input position is randomly selected to be $>1^{\prime\prime}.0$ away from $\ge3\sigma$ sources.
We then repeated the same procedures for the source extraction using the artificial map as described in \S 3.1.

The completeness is measured as the recovery rate of the injected model sources; when an input source is detected within 1$^{\prime\prime}$.0 of the injected position with $\ge 4\sigma$, the source is considered to be recovered.
Model sources with flux densities over the range $S=$0.02--0.6~mJy in steps of 0.02~mJy are considered and 
the procedures are repeated 1000 times for each flux density bin ($\Delta S=$0.02~mJy).
The results are summarized in Fig.~\ref{f4} as a function of input flux density.
The completeness of our source extraction is $\sim70$\% at SNR=4 and rises to $\sim90$\% at SNR=5.
The completeness reaches $\sim$100\% at SNR$\gtrsim$6.0.
We consider the completeness in calculating source counts in \S 4.3, although it does not matter significantly for $>5\sigma$ sources.

In the course of the completeness measurement, we also evaluate the effect of flux boosting.
It is known that the flux density of sources in a signal-to-noise limited catalog tend to be boosted as a whole due to random noise fluctuations and the shape of the source count distribution (e.g., \citealt{1998PASP..110..727H}; \citealt{2002MNRAS.331..817S}; \citealt{2006MNRAS.372.1621C}).
We plot the ratio of input and output flux density against the input flux density in Fig.~\ref{f4}.
The results show that the effect is negligible if we consider only those sources detected at $>5\sigma$.
Hence we do not consider the flux boosting effect in the following discussion.

\subsection{Source Catalog}

\begin{center}
\begin{deluxetable*}{ccccccccc}
\tabletypesize{\scriptsize}
\tablecaption{ADF22 Source Properties}
\tablewidth{0pt}
\tablehead{
\colhead{(1)}  & \colhead{(2)} &\colhead{(3)} & \colhead{(4)}  & \colhead{(5)} & \colhead{(6)}   & \colhead{(7)} & \colhead{(8)} &  \colhead{(9)}\\
\colhead{ALMA ID}  & \colhead{ALMA NAME} & \colhead{ID (U15)} &\colhead{AzTEC ID} & \colhead{$S_{\rm AzTEC}$}  & \colhead{$\sigma_{\rm ALMA}$}   & \colhead{SNR$^{\rm pk}_{\rm ALMA}$} & \colhead{$S_{\rm ALMA}$} &  \colhead{$z_{\rm spec}$}\\
 & & & (mJy) & (J2000) & ($\mu$Jy beam$^{-1}$)   &  &(mJy) &  
}
\startdata
ADF22.1*  & ALMAJ221732.41+001743.8  &  ADF22a &  AzTEC1  & 11.3$^{+0.9}_{-0.7}$ &  72  &  58.1   &   5.60 $\pm$ 0.13   & 3.092$^{a}$ \\ 
ADF22.2  & ALMAJ221736.11+001736.7 &  --- &  --- & ---   &  63  &  31.8   &   2.02 $\pm$ 0.02   & ...   \\ 
ADF22.3  & ALMAJ221735.15+001537.2 &  ADF22b &  AzTEC77  & 2.4$^{+0.9}_{-0.8}$  &  66  &  27.0   &   1.89 $\pm$ 0.04   & 3.096$^{d}$  \\
ADF22.4  & ALMAJ221736.96+001820.7  & --- &  AzTEC14  & 4.5$^{+0.8}_{-0.8}$ &   72  &  26.6   &   1.95 $\pm$ 0.05   & 3.091$^{b}$ \\ 
ADF22.5  &  ALMAJ221731.48+001758.0  &  --- &  ---  &--- &  99  &  20.3   &   2.43 $\pm$ 0.20   & ...   \\ 
ADF22.6  &  ALMAJ221735.83+001559.0  &ADF22c &  ---  &--- &    69  &  19.1   &   1.45 $\pm$ 0.09   & 3.089$^{e}$ \\ 
ADF22.7  &  ALMAJ221732.20+001735.6  &ADF22i &  AzTEC1  & 11.3$^{+0.9}_{-0.7}$ &   86  &  18.7   &   1.65 $\pm$ 0.07   & 3.097$^{c}$ \\
ADF22.8  &  ALMAJ221737.11+001712.3  & ADF22d &  ---  &--- &    77  &  15.0   &   1.19 $\pm$ 0.06   & 3.090$^{f}$ \\ 
ADF22.9  & ALMAJ221736.54+001622.6  & ADF22e &  ---  &--- &    60  &  12.8   &   0.82 $\pm$ 0.08   & 3.095$^{f}$ \\
ADF22.10  & ALMAJ221737.10+001826.8  &  --- &  AzTEC14  & 4.5$^{+0.8}_{-0.8}$ &   71  &  9.8   &   0.72 $\pm$ 0.04   & ...   \\
ADF22.11  &  ALMAJ221737.05+001822.3  & ADF22f &  AzTEC14  & 4.5$^{+0.8}_{-0.8}$ &  77  &  9.5   &   0.79 $\pm$ 0.05   & 3.093$^{f}$ \\ 
ADF22.12  & ALMAJ221732.00+001655.4  & ADF22g &  ---  &--- &   71  &  8.8   &   0.63 $\pm$ 0.03   & 3.091$^{f}$ \\
ADF22.13  & ALMAJ221737.42+001732.4  & --- &  ---  &--- &   81  &  8.0   &   0.79 $\pm$ 0.05   & ...   \\ 
ADF22.14  & ALMAJ221731.34+001639.6  & --- &  ---  &--- &   99  &  7.5   &   0.98 $\pm$ 0.13   & ...   \\ 
ADF22.15  & ALMAJ221732.77+001727.5  & --- &  ---  &--- &   79  &  5.3   &   0.50 $\pm$ 0.08   & ...   \\ 
ADF22.16  & ALMAJ221736.81+001818.0  &  ADF22h &  AzTEC14  & 4.5$^{+0.8}_{-0.8}$ &   77  &  5.3   &   0.56 $\pm$ 0.07   & 3.085$^{f}$ \\ 
ADF22.17  & ALMAJ221737.69+001814.4  & --- &  AzTEC14  & 4.5$^{+0.8}_{-0.8}$ &  66  &  5.1   &   0.60 $\pm$ 0.09   & ...   \\ 
ADF22.18** & ALMAJ221732.23+001527.8  &  --- &  ---  & --- &   82  &  5.3  &   0.44 $\pm$ 0.05   & 2.105$^{g}$
\enddata 
\tablecomments{
\newline
$(1)$ ID in this paper; 
\newline
$(2)$ Source name which represents the coordinate in the wcs system;
\newline
$(3)$ ID defined in \citet{2015ApJ...815L...8U}; 
\newline
$(4)$ ID of AzTEC source in \citet{2014MNRAS.440.3462U}. We list it if an ALMA SMG is located within the 30$^{\prime\prime}$ AzTEC beam;
\newline
$(5)$ Flux density of AzTEC source (\citealt{2014MNRAS.440.3462U});
\newline
$(6)$ 1$\sigma$ sensitivity at a given source position after the primary beam correction;
\newline
$(7)$ The signal to noise ratio, which is defined as a ratio of peak flux density over the 1$\sigma$ sensitivity;
\newline
$(8)$ The flux density and uncertainty measured by the {\sc casa} task, {\sc imfit};
\newline
$(9)$ Spectroscopic redshift ($z_{\rm spec}$) if a give source have it. Lines to be used to determine $z_{\rm spec}$ and references are follows: 
a; $^{12}$CO(3-2) (Yun et al. in preparation), 
b; $^{12}$CO(9-8) (Umehata et al. in preparation; Hayatsu et al. in preparation), 
c; \textsc{[Cii]} 157.7$\mu$m (Umehata et al. in preparation), 
d; $^{12}$CO(3-2) (\citealt{2013MNRAS.429.3047B}, see also \citealt{2005MNRAS.359.1165G}, \citealt{2005ApJ...622..772C})
e; Ly$\alpha$ (\citealt{2005ApJ...622..772C}), 
f; \textsc{[Oiii]}$\lambda5007$ (\citealt{2015ApJ...799...38K}, \citeyear{2016MNRAS.455.3333K}),
g; Ly$\alpha$ (\citealt{2004ApJ...614..671C}).
\newline
* This source was identified as a primary counterpart of SSA22-AzTEC1 using Submillimeter Array (SMA) in \citet{2010ApJ...724.1270T}.
\newline
** The properties of ADF22.18 are measured using the DEEP/HIRES map.
}
\end{deluxetable*}
\end{center}

\begin{figure*}
\epsscale{1.05}
\plotone{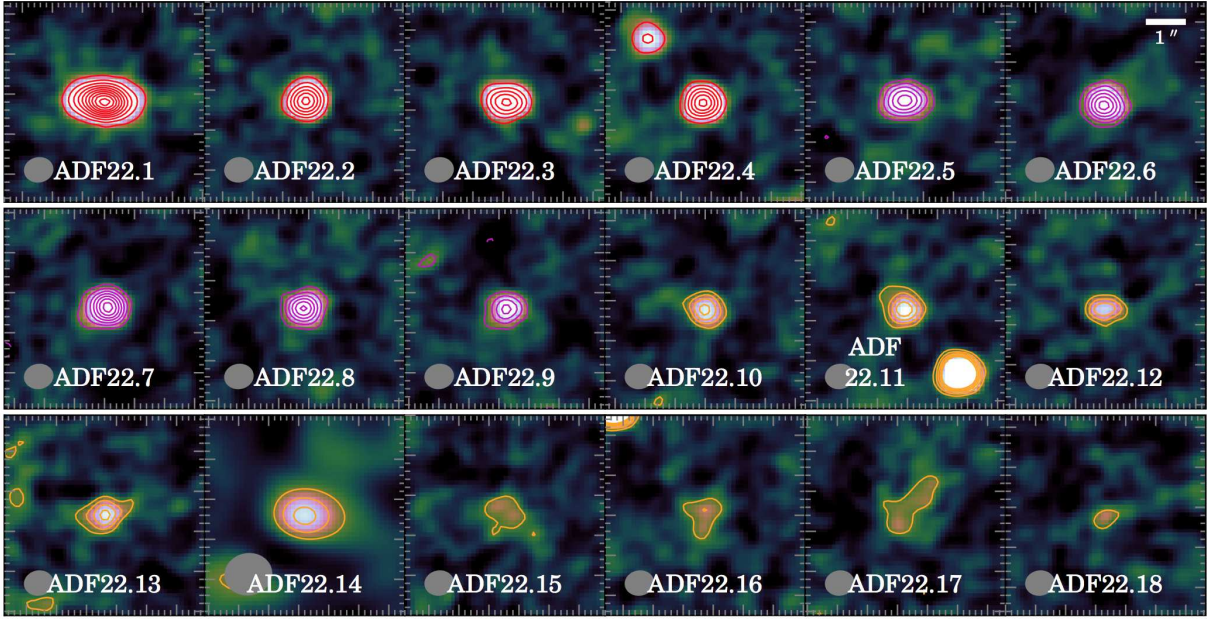}
\caption{
ALMA 1.1~mm continuum maps of 18 sources in ADF22.
Each map is $5^{\prime\prime}\times5^{\prime\prime}$ in size.
Generally we show the DEEP/HIRES map to display the better angular resolution image.
We use the FULL/LORES map only for ADF22.14, which are not covered by the DEEP/HIRES map (see Fig.~\ref{f2} for the source distributions and covered area of the FULL/LORES and DEEP/HIRES map).
IDs are shown in the bottom in the maps and synthesized beam are displayed in the bottom left of each map.
The 1.1~mm ALMA contours are in steps of 5$\sigma$ starting at $\pm5\sigma$ (red, ADF22.1--ADF22.4), in steps of 3$\sigma$ starting at $\pm3\sigma$ (magenta, ADF22.5--ADF22.9), and in steps of 2$\sigma$ starting at $\pm3\sigma$ (orange, ADF22.10--ADF22.18).
}
\label{stamp}
\end{figure*}

The measured properties of the ADF22 SMGs are summarized in Table 2.
The first column lists the source IDs in this paper, which are generally ranked in terms of SNR.
The nine SMGs reported in \citet{2015ApJ...815L...8U} are noted in the second column.
There are three AzTEC sources in our field (\citealt{2014MNRAS.440.3462U}).
We listed the IDs and flux densities of the corresponding AzTEC source if a given ALMA SMG is located within the AzTEC beam (FWHM; $d=30^{\prime\prime}$).
We give the flux density, measured by Gaussian fitting with {\sc casa}, {\sc imfit}.
We note that these measurements are from the FULL/LORES map except for ADF22.18, which was detected at $>5\sigma$ only in the DEEP/HIRES map. 

We also report the spectroscopic redshift ($z_{\rm spec}$), if known.
Eleven of the 18 SMGs have $z_{\rm spec}$, ten of which have $z\simeq3.09$.
In addition to eight $z\simeq3.09$ SMGs reported in \citet{2015ApJ...815L...8U}, we adopt $z_{\rm spec}$ for three SMGs, ADF22.4, ADF22.7, and ADF22.18.
The redshifts of ADF22.4 and ADF22.7 is determined from our recent ALMA observations ($^{12}$CO(9-8) at $z=3.091$ and \textsc{[Cii]} 157.7~$\mu$m at $z=3.097$, respectively; Umehata et al., in preparation).
ADF22.18 coincides with a $z=2.015$ radio source, RG J221732.22+001528.2, reported in \citet{2004ApJ...614..671C} with a small projected separation ($0^{\prime\prime}.4$).
We show the ALMA 1.1~mm images of the 18 SMGs in Fig.~\ref{stamp}.

One reasonable interest regarding the ALMA source catalog is the overlap between the previously known galaxy populations and our newly discovered ALMA SMGs.
Within the survey area of the FULL/LORES map (Fig. \ref{f2}), there are 19 LAEs (\citealt{2005ApJ...634L.125M}; \citealt{2013ApJ...765...47N}; \citealt{2014ApJ...795...33E}), five LBGs (\citealt{2003ApJ...592..728S}; \citealt{2013ApJ...765...47N}; \citealt{2014ApJ...795...33E}), and ten $K$-band selected galaxies (\citealt{2015ApJ...799...38K}; \citealt{2016MNRAS.455.3333K}) within the protocluster (i.e., with $z_{\rm spec}=3.06-3.12$; \citealt{2004AJ....128.2073H}; \citealt{2005ApJ...634L.125M}).
None of the ALMA SMGs (including candidates in the supplementary source catalog) have LAE/LBG counterparts, which shows that these rest-frame UV selected galaxies are clearly separated populations compared to galaxies individually detected by ALMA.
In contrast, five out of ten K-band selected galaxies are securely detected by ALMA.
Therefore such a relatively massive galaxy population (stellar mass, $\gtrsim10^{10-11}M_\odot$; \citealt{2013ApJ...778..170K}) selected at rest-frame optical wavelengths appears to significantly overlap with the ALMA population.
This trend is broadly consistent with recent works in SXDF (\citealt{2015ApJ...811L...3T}) and HUDF (\citealt{2016arXiv160600227D}; \citealt{2016arXiv160706769A}) as well as previous studies of ALMA SMGs (\citealt{2014ApJ...788..125S}).
We summarize the relationship to other populations in Table 6 in Appendix B.

\section{Discussion}

\subsection{Resolving the AzTEC map with ALMA}

\begin{figure}
\epsscale{1.15}
\plotone{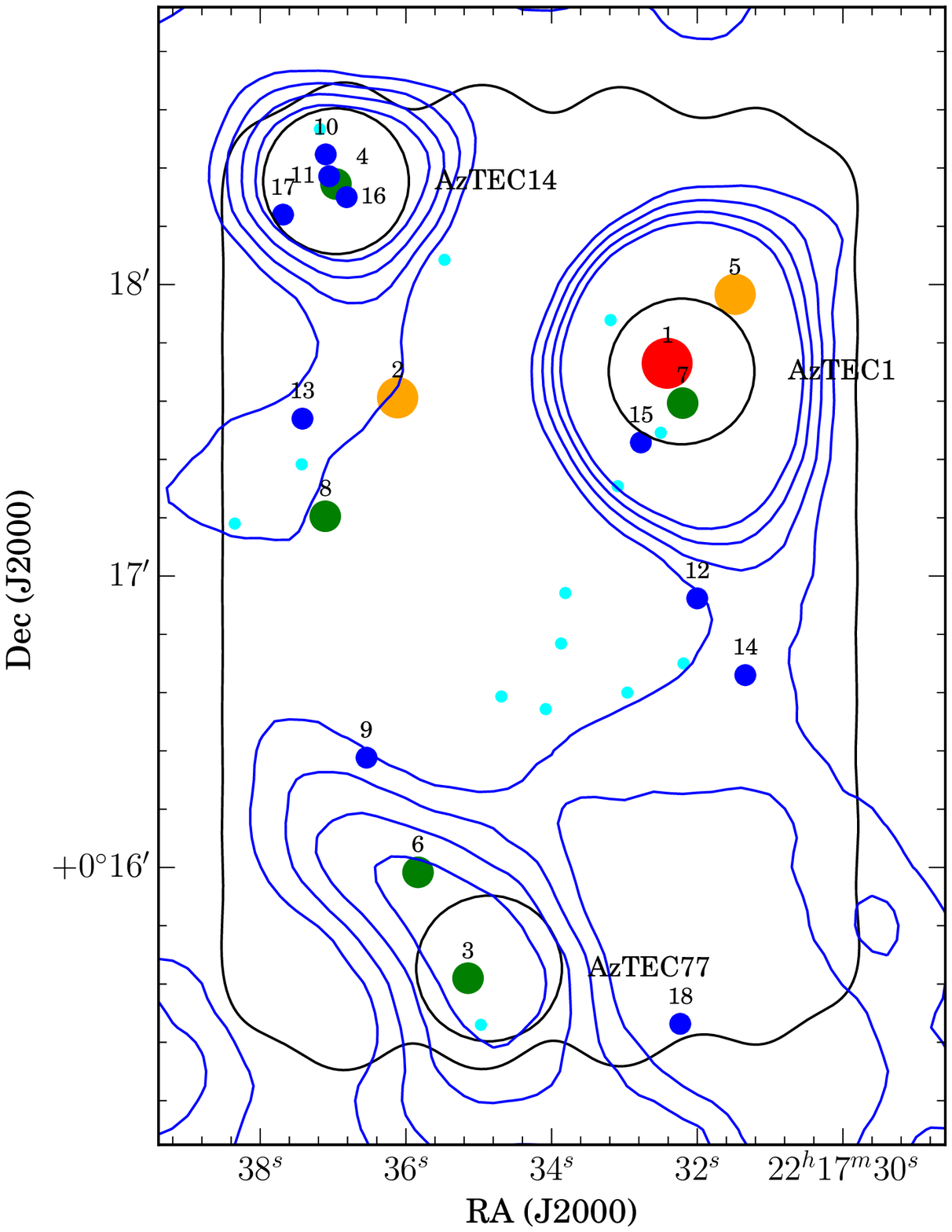}
\caption{
The distributions of the ALMA SMGs and AzTEC sources in the ADF22.
Filled circles represent the ALMA SMGs, the size and color of which stand for their flux density (red: $S_{\rm 1.1mm}>5$ mJy, orange: 2 mJy $<S_{\rm 1.1mm}\le5$ mJy, green: 1 mJy $<S_{\rm 1.1mm}\le2$ mJy, blue: $S_{\rm 1.1mm}\le 1.0$ mJy). 
The associated number is the ALMA source ID in this paper.
Filled cyan circles represent the positions of ALMA sources in the supplementary catalog.
The positions of three AzTEC sources are shown using large black circles, the diameters of which is equivalent to the beam FWHM of the AzTEC map ($d=30^{\prime\prime}$).
We also plot 0, 1, 2, and 3 $\sigma$ contours of the AzTEC 1.1~mm map (1$\sigma$ = 0.7~mJy; \citealt{2014MNRAS.440.3462U}).
The black contour is the 50\% coverage area of the FULL/LORES map, which is same as Fig.~\ref{f2}.
The ALMA mosaic revealed the source distribution below the AzTEC detection limit across the area.
}
\label{dist_alma}
\end{figure}

Several hundred submm/mm sources discovered by single-dish telescopes have been observed by submm/mm interferometers to date.
Some appear to resolve into multiple, individual SMGs while others have a unique counterpart above a given detection limit (e.g., \citealt{2000MNRAS.316L..51G}; \citealt{2006ApJ...640..228T}; \citealt{2008MNRAS.387..707Y}; \citealt{2011ApJ...726L..18W}; \citealt{2012ApJ...761...89B}; \citealt{2012A&A...548A...4S}; \citealt{2013ApJ...776...22H}; \citealt{2015A&A...577A..29M}; \citealt{2015ApJ...807..128S}).
For instance, \citet{2012ApJ...761...89B} reported that, using the SMA, three of 16 SCUBA sources are composed of multiple objects.
From ALMA observations,
\citet{2013ApJ...768...91H} and \citet{2015ApJ...807..128S} suggested that $\sim$30--40 \% of LABOCA or bright SCUBA2 sources (with flux densities of $S_{\rm 870\mu m}=4-15$~mJy and $S_{\rm 850\mu m}\sim8-16$~mJy, respectively) are resolved into multiple SMGs brighter than $S_{\rm 870\mu m}\sim$1mJy, making most of components ``ULIRGs''.
These results suggest that the relatively poor angular resolution of single-dish imaging ($\gtrsim15^{\prime\prime}$) causes significant source blending  and complicate our interpretation of the nature of SMGs.
Now ALMA enables us to resolve not only individual single-dish sources but also relatively faint sources spread over a wider area.

\subsubsection{Comparison of the AzTEC and ALMA map}

The AzTEC/ASTE survey of SSA22 by \citet{2009Natur.459...61T} and \citet{2014MNRAS.440.3462U} presented a 1.1~mm image of ADF22.
The 30$^{\prime\prime}$ resolution image has a 1$\sigma$ depth of 0.7~mJy beam$^{-1}$, and detects three sources above a 3.5$\sigma$ detection threshold within ADF22 (SSA22-AzTEC1, SSA22-AzTEC14, and SSA22-AzTEC77; hereafter AzTEC1, AzTEC14, and AzTEC77, respectively).
Two of the three AzTEC sources, AzTEC1 and AzTEC14, have two and five ALMA SMGs located within the FWHM of the AzTEC beam, respectively. In contrast AzTEC77 has only one associated ALMA SMG (Fig.~\ref{dist_alma})\footnote{AzTEC77 shows an elongated profile, which should be caused by ADF22.6, as illustrated in Fig.~\ref{dist_alma}. Therefore the profile itself should reflect the two ALMA SMGs. However, ALMA6 is located at outside of the AzTEC beam of AzTEC77 and so does not contribute to the measured flux of AzTEC77 significantly. Hence here we treat ADF22.3 as an unique counterpart of AzTEC77.}.
In summary, three AzTEC sources are resolved into eight ALMA SMGs.
The result is in line with previous ALMA studies following up single-dish submm/mm sources with interferometers, which reports that a significant fraction of submm/mm sources detected by single-dish telescopes are found to be intrinsically multiple SMGs (e.g., \citealt{2013ApJ...776...22H})

\begin{figure}
\epsscale{1.15}
\plotone{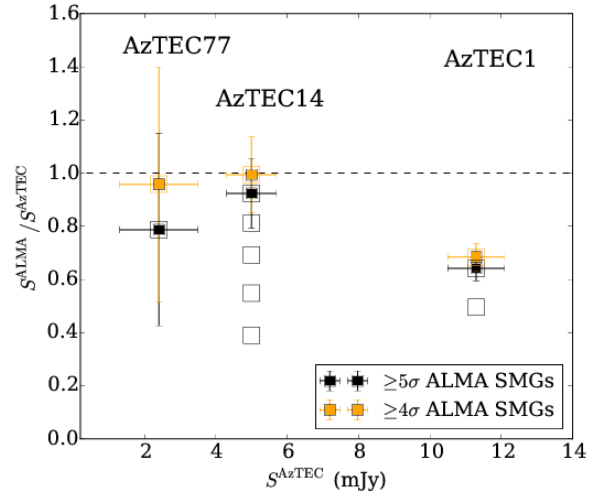}
\caption{
A comparison of the flux measurements in the AzTEC and ALMA maps for the three AzTEC sources; AzTEC1, AzTEC14, and AzTEC77.
$S^{\rm AzTEC}$ shows the flux density of the AzTEC sources and $S^{\rm ALMA}$ stands for the integration of that of the ALMA SMGs within the AzTEC beam.
The filled black squares show the flux ratio,  $S^{\rm ALMA}$/$S^{\rm AzTEC}$, as a function of $S^{\rm AzTEC}$, considering  ALMA SMGs with $\ge5\sigma$.
The filled orange squares show the flux ratio, including the sources in the supplementary catalog.
Empty squares represent the ratio in the case of cumulative flux density of individual ALMA SMGs, which are in order of flux density.
While $S^{\rm AzTEC}$ and $S^{\rm ALMA}$ match within the range of error for AzTEC14 and AzTEC77, 
$S_{\rm ALMA}$ is only 64$\pm5$\% of $S^{\rm AzTEC}$ in the case of AzTEC1. 
}
\label{flux_comp}
\end{figure}

To check the relative flux scales,
we compared the flux density between the AzTEC sources and the associated ALMA SMGs.
For the AzTEC sources, deboosted flux densities in \citet{2014MNRAS.440.3462U} ($S^{\rm AzTEC}$) are considered here.
Regarding the measurements for the ALMA SMGs, 
we calculate the sum of the flux densities of the SMGs with $\ge5\sigma$ within the FWHM of the AzTEC beam ($S^{\rm ALMA}$). 
Fig.~\ref{flux_comp} shows the results for the three AzTEC sources.
The ratio between the two, $S^{\rm ALMA}$/$S^{\rm AzTEC}$, is in agreement within error bars for AzTEC77 and AzTEC14, which suggests that the ALMA SMGs discovered from this survey account for the majority of the flux density of the AzTEC sources.
However, the situation is not the same for AzTEC1, the brightest AzTEC source in the SSA22 field, which has $S^{\rm ALMA}$/$S^{\rm AzTEC}=0.64\pm0.05$.
The discrepancy could imply the existence of additional components of 1.1~mm emission not accounting for in the ALMA map.
One possibility is that we are missing a number of fainter and/or diffuse sources (e.g., \citealt{2013ApJ...776...22H}; \citealt{2015ApJ...807..128S}).
If we include sources in the supplementary catalog with SNR=$4-5$, then the flux ratio $S^{\rm ALMA}$/$S^{\rm AzTEC}$ 
gets closer to unity for all three AzTEC sources (Fig.~\ref{flux_comp}).
But these sources is still insufficient to explain the case of AzTEC1.
Much fainter sources might also contribute to the AzTEC sources for AzTEC1.

\citet{2010ApJ...724.1270T} gave us another clue.
Tamura et al. reported the 860$\mu$m flux density of ADF22.1, $S_{\rm 860\mu m}=12.2\pm2.3$~mJy, using the Submillimeter Array\footnote{\bf \citet{2010ApJ...724.1270T} found only one source toward SSA22-AzTEC1 and ADF22.7 was not detected significantly in the paper.}. They  obtained the natural-weighted synthesized beam, $3^{\prime\prime}.43\times1^{\prime\prime}.92$ (P.A. 34.3 deg), and found that the source was likely not to be resolved.
If we predict 1.1~mm flux density of ADF22.1 using an averaged SMG template from the ALESS survey (\citealt{2014MNRAS.438.1267S}) scaled to the 860~$\mu$m flux density, the expected value is $S_{\rm 1.1mm}\approx$6.0~mJy.
The estimate is broadly consistent with our measurement, which suggests that we are measureing the vast majority of the dust emission from ADF22.1.

There are two ALMA SMGs, ADF22.5 and ADF22.15, located just outside of the beam for AzTEC1 (Fig.~\ref{dist_alma}).
If we take into account ADF22.5 and ADF22.15, the integral of the flux densities of the ALMA SMGs approach the flux density of AzTEC1 ($S^{\rm ALMA}$/$S^{\rm AzTEC}=0.94\pm0.07$).
In order to investigate whether these SMGs indeed contribute to the flux density of AzTEC1 in the AzTEC map,
we made a ``model'' AzTEC image convolving the ALMA map with a Gaussian kernel with a FWHM of 30$^{\prime\prime}$.
AzTEC1 has flux density 7.1~mJy in the model map, which is consistent with the sum of flux density of the ALMA SMGs within the beam for AzTEC1, not including ADF22.5 and ADF22.15.
Therefore these two nearby SMGs are likely not to account for the discrepancy.
We do not consider a systematic error on the absolute flux accuracy for the AzTEC map throughout the above discussion.
Although it is difficult to estimate the influence for our small sample, ALMA surveys of a significant number of AzTEC sources will allow us to examine the effect.

As displayed in Fig.~\ref{dist_alma}, we compared the spatial distribution of SMGs discovered by ALMA and the contours of the AzTEC 1.1~mm emission, below the AzTEC detection limit (3.5$\sigma$).
Our ALMA map has discovered ten SMGs outside the AzTEC source positions.
We found that nine of the ten ALMA SMGs are located within or in the vicinity of the area where the 1.1~mm flux density is 0~--~2.1~mJy~beam$^{-1}$ in the AzTEC map. 
This may show that the structure traced by the faint AzTEC emission, below 3$\sigma$, reflects the distribution of 1.1mm sources to a certain degree.

As we reported in \citet{2014MNRAS.440.3462U}, the AzTEC catalog is significantly incomplete at around 3.5$\sigma$ (i.e., the detection threshold). The completeness of the AzTEC map is only $\sim50$\% at 2~mJy.
Among the ADF22 sources with flux density $\sim$2~mJy, two SMGs, ADF22.2 and ADF22.5, are found to be located outside of the AzTEC beam. Our results therefore show that ALMA mosaic has capability to find such bright SMGs which was missed by previous submm/mm surveys taken with single-dish telescopes.



\subsubsection{The origin of multiplicity in a overdense environment}

There has been a debate on whether multiple individual SMGs of sources which are identified with a single-dish telescope are physically connected or instead are just a line-of sight projection (e.g., \citealt{2007MNRAS.380..199I}; \citealt{2011ApJ...726L..18W}; \citealt{2013ApJ...776...22H}; \citealt{2013MNRAS.434.2572H}; \citealt{2015ApJ...807..128S}; \citealt{2015MNRAS.446.1784C}).
For one source, \citet{2011ApJ...726L..18W} proposed that it was composed of multiple, physically unrelated SMGs.
In contrast, \citet{2015ApJ...811L...3T} showed that one AzTEC source fragments into two H$\alpha$ emitters at $z=2.53$, which supports the physical connection between the two (see also \citealt{2016arXiv160702331Y}).
\citet{2015ApJ...807..128S} suggested that a portion of bright submm sources arise from physically related SMGs since the number of detected ALMA SMGs in the vicinity of the SCUBA2 sources are two orders of magnitude higher than the general field.
On the other hand, recently \citet{2015MNRAS.446.1784C} claimed that physically unrelated SMGs can also reproduce the single-dish sources in their semi-analytic model (see also \citealt{2013MNRAS.434.2572H}).

Among 18 SMGs in ADF22, 11 SMGs have $z_{\rm spec}$ (\citealt{2015ApJ...815L...8U}; in prep), 10 of which are at $z= 3.085 - 3.097$ (Table 2).
All of the brightest ALMA counterparts of the AzTEC sources are at $z=3.09$ and AzTEC1 and AzTEC14 have multiple ALMA SMGs at $z\simeq3.09$ within the AzTEC beam ($d\sim230$ kpc at $z=3$).
These results favor that physically connected multiple SMGs appear as a single single-dish source, at least in dense environment such as center of a protocluster.
The overdensity of galaxies on such scales is conductive to mergers and dissipative interactions (e.g., \citealt{2012ApJ...752...39T}) and 
hence such small scale over-densities of SMGs may indicate that SMGs undergo merger-induced star formation.
Furthermore, there are other $z=3.09$ ALMA SMGs outside the AzTEC beam across the entire field of ADF22 (Fig.~\ref{dist_alma}, Table 2).
This suggests that intense dusty star-formation may also be enhanced by the environment on a large-scale (\citealt{2015ApJ...815L...8U}; see also \citealt{2004ApJ...611..725B}; \citealt{2009Natur.459...61T}; \citealt{2014MNRAS.440.3462U}; \citealt{2015ApJ...808L..33C}; \citealt{2016arXiv160507176H}; \citealt{2016arXiv160304437C}).

\subsection{Size measurement and star formation rate surface density}

\begin{figure}
\epsscale{1.15}
\plotone{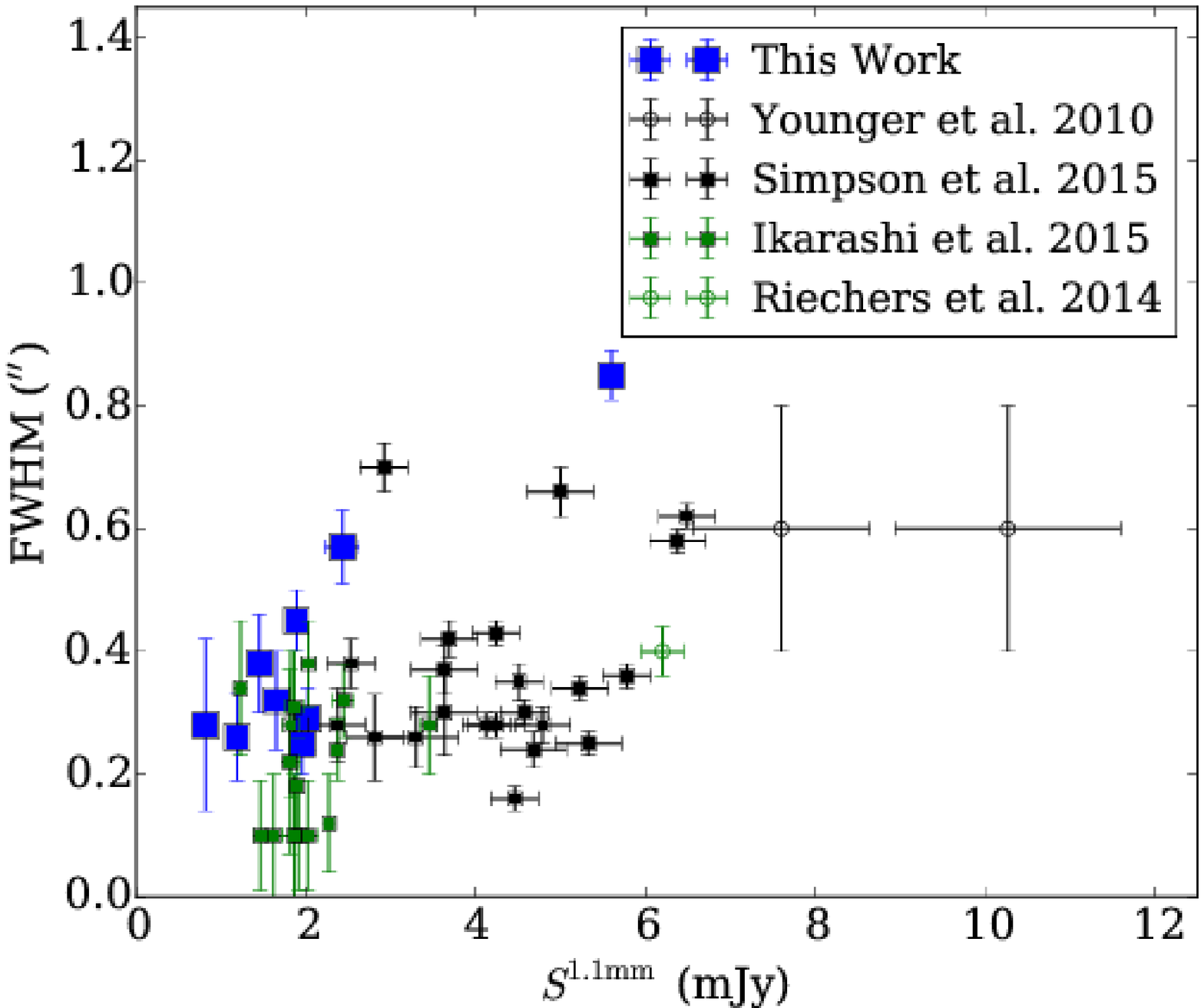}
\caption{
Angular size distribution of the 1.1~mm emission from SMGs as a function of their 1.1~mm flux density.
The horizontal axis shows a deconvolved FWHM of the major axis from two-dimensional Gaussian fitting
The results of ADF22 sources are shown.
We also shows those of previous works (\citealt{2010MNRAS.407.1268Y}; \citealt{2014ApJ...796...84R}; \citealt{2015ApJ...799...81S}; \citealt{2015ApJ...810..133I}).
The flux densities at 870 $\mu$m (\citealt{2015ApJ...799...81S}) and  at 890 $\mu$m (\citealt{2010MNRAS.407.1268Y}) are scaled to 1.1~mm flux density using $S_{\rm 1.1~mm}$/$S_{\rm 870\mu m}$ = 0.56 and $S_{\rm 1.1~mm}$/$S_{\rm 890\mu m}$ = 0.58, respectively. We assume a modified black body with typical values for SMGs (spectral index of $\beta$ = 1.5, dust temperature of 35 K (e.g., \citealt{2014MNRAS.438.1267S}), and $z = 2.5$). 
The source sizes are generally consistent with the works and there is no clear environmental dependence.
}
\label{size}
\end{figure}


Two-dimensional Gaussian fits to the ALMA SMGs in the image plane using the DEEP/HIRES map suggest that ADF22 sources are generally resolved.
The results for the nine brightest SMGs with SNR$>10$ are summarized in Table 3.
The deconvolved major axes are 0$^{\prime\prime}$.25 -- 0$^{\prime\prime}$.85 and the median value for the nine SMGs is 
$0^{\prime\prime}.32^{+0^{\prime\prime}.13}_{-0^{\prime\prime}.06}$ (Gaussian FWHM; $2.4^{+1.0}_{-0.4}$ physical kpc at $z=3.09$).
Seven of the nine SMGs have $z_{\rm spec}\simeq3.09$, which leads the almost same median value,  $0^{\prime\prime}.32^{+0^{\prime\prime}.13}_{-0^{\prime\prime}.07}$ if we only consider such robust protocluster members.
The size distribution in ADF22 is generally consistent with previous measurements for SMGs in other fields (Fig.~\ref{size}).
\citet{2015ApJ...799...81S} measured the size for 23 bright SMGs observed at 870 $\mu$m and derived a median angular size FWHM=$0^{\prime\prime}.30\pm0^{\prime\prime}.04$ and 
\citet{2015ApJ...810..133I} reported that 13 SMGs, which were observed at 1.1~mm and thought to be at $z\sim3-6$, have  a median FWHM of $0^{\prime\prime}.20^{+0^{\prime\prime}.03}_{-0^{\prime\prime}.05}$.
These results suggest that dusty star formation occurs in compact regions, a few kiloparsecs in extent.
There does not appear to be a significant difference between the size of SMGs within the $z=3.09$ protocluster and other SMGs.
This might suggest that the local mechanism triggering intense starbursts does not significantly depend on the large-scale environment.

The radio continuum also provides a tool to measure the scales of dusty starburst galaxies like SMGs (e.g., \citealt{2004ApJ...611..732C}; \citealt{2008MNRAS.385..893B}; \citealt{2016arXiv160707710R}).
These measurements shows relatively extended profile (e.g., a median of 5~kpc; \citealt{2008MNRAS.385..893B}), which is larger than the bright ALMA SMGs discussed above. Up to date some explanation has been proposed.
For instance, \citet{2016arXiv160707710R} suggests that the relatively compact star formation is seen in SMGs with higher star-formation rate (SFR) while main-sequence star-forming galaxies have different scales.
\citet{2015ApJ...799...81S} indicates that the difference between the diffusion length of cosmic rays and far-infrared photons can account for it.
Deep and high angular resolution radio imaging in ADF22 in the future will provide us with important clue to clarify the origin of the discrepancy between the radio and FIR sizes of SMGs.


The flux density and measured source size allow us to derive star-formation rate surface density ($\Sigma_{\rm SFR}$) and investigate the condition of star-formation.
We estimated $L_{\rm IR}$ [8-1000 $\mu$m] using SED templates of well-studied starburst galaxies (\citealt{1998ApJ...509..103S}; \citealt{2010Natur.464..733S}; \citeyear{2014MNRAS.438.1267S}) in the same manner as in \citet{2015ApJ...815L...8U}.
The SFR is in turn derived from $L_{\rm IR}$ using the empirical calibration by \citet{1998ARA&A..36..189K} adjusted to the Kroupa initial mass function (\citealt{2001MNRAS.322..231K}).
Since ADF22.2 and ADF22.5 do not have $z_{\rm spec}$, we assume $z=3.0$ (a median value for AzTEC sources; \citealt{2012ApJS..200...10S}).
Finally $\Sigma_{\rm SFR}$ was calculated using the size of the 1.1~mm continuum emission and the SFR divided by a factor of two, following \citet{2015ApJ...799...81S} (Table 3).
The median value is $\Sigma_{\rm SFR}$=50 $M_\odot$ yr$^{-1}$ kpc$^{-2}$, significantly lower than the predicted Eddington limit for radiation pressure supported disks (e.g., \citealt{1999ApJ...517..103E}; \citealt{2005ApJ...630..167T}; \citealt{2010MNRAS.407.1268Y}; \citealt{2011ApJ...727...97A}; \citealt{2013Natur.496..329R}; \citealt{2014ApJ...796...84R}).
Therefore starbursts seen in bright SMGs in ADF22 are not Eddington limited as a whole.
As some authors have noted (e.g., \citealt{2015ApJ...799...81S}), if SMGs have clumpy structure, the individual components might be Eddington limited, as has been claimed from some recent high angular resolution ALMA images of the brightest SMGs (e.g., Hatsukade et al. 2015; Iono et al. 2016, but see also \citealt{2016arXiv160909649H}).
Further observations capable of resolving sub-kpc structures ($\lesssim 0^{\prime\prime}.1$ angular resolution) are required to probe such a scenario  for the SMGs in ADF22.







\subsection{Number Counts}

\begin{figure*}
\epsscale{1.15}
\plotone{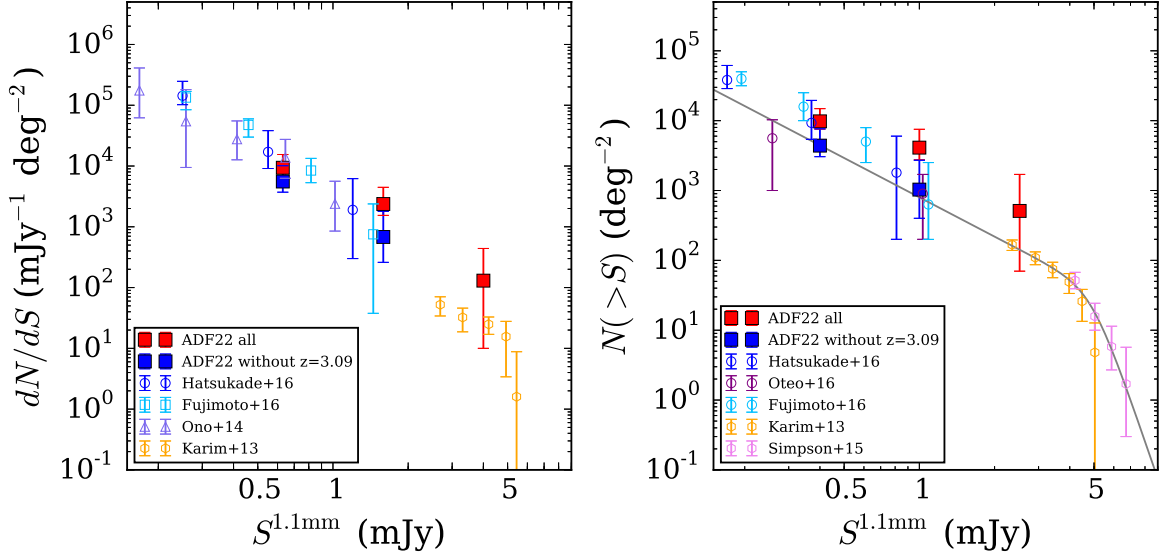}
\vspace{-6.5cm}
\caption{
Differential (left panel) and cumulative (right panel) number counts in ADF22 and other fields at 1.1~mm.
The counts in ADF22 are calculated in two ways.
We show the counts calculated from our full sample of the 17 SMGs above $5\sigma$ in the FULL/LORES map.
We also show the counts using 7 SMGs which don't have $z_{\rm spec}=3.09$.
Previous ALMA results (\citealt{2013MNRAS.432....2K}; \citealt{2015ApJ...807..128S}; \citealt{2016ApJ...822...36O}; \citealt{2014ApJ...795....5O}; \citealt{2016ApJS..222....1F}; \citealt{2016PASJ...68...36H}) are also shown. The flux density of the counts are scaled to 1.1~mm flux density assuming a modified black body (similar with Fig.~\ref{size}; $S_{\rm 1.1~mm}$/$S_{\rm 870\mu m}$ = 0.56, $S_{\rm 1.1~mm}$/$S_{\rm 1.2~mm}=1.29$, $S_{\rm 1.1~mm}$/$S_{\rm 1.3~mm}=1.48$).
The gray curve is the best-fit functions of a double-power law for the counts from our 7 SMGs without $z=3.09$, \citet{2013MNRAS.432....2K}, and \citet{2015ApJ...807..128S}.
The cumulative counts from 17 SMGs shows about five times excess at $S_{\rm 1.1~mm}\gtrsim1.0$~mJy, which should be caused by the $z=3.09$ protocluster.
}
\label{count}
\end{figure*}

In this section, we report the 1.1~mm number counts in ADF22, which is one of the most fundamental parameters in describing the evolution of this population. 
Previously there have been a number of studies investigating submm/mm number counts from a variety of ALMA data sets; LABOCA/SCUBA2 source follow-up (\citealt{2013MNRAS.432....2K}; \citealt{2015ApJ...807..128S}), the use of calibration field (\citealt{2016ApJ...822...36O}), wide range of archival data (\citealt{2013ApJ...769L..27H}; \citealt{2014ApJ...795....5O}; \citealt{2015A&A...584A..78C}; \citealt{2016ApJS..222....1F}), and contiguous mapping (\citealt{2016PASJ...68...36H}; \citealt{2016arXiv160600227D}; \citealt{2016arXiv160706769A}).
Here we present the number counts in a $\sim$7 arcmin$^2$ contiguous field in the remarkable high density region in the early universe, 
well-suited for investigating the environmental dependence of SMG source counts.

To minimize the number of contaminants by false positives, we utilized only the 17 SMGs detected with SNR $> 5\sigma$ in the FULL/LORES map. 
We excluded ADF22.18 since the SNR of this source is $<5$ in the FULL/LORES map. 
To investigate how the existence of the $z=3.09$ protocluster affects the counts, we calculated the counts in two ways.
First we included all 17 SMGs.
Second, we exclude the ten SMGs with known $z_{\rm spec}=3.09$ (Table 2) and used only the remaining seven SMGs.
Flux boosting effects were not considered (see \S3.2).
We corrected for the completeness of the source extraction, although the counts should not be significantly affected by the correction.
We calculated the number counts and associated errors in the same way as \citet{2016PASJ...68...36H}.
Fig.~\ref{count} shows the differential and cumulative counts (in the left and right panels, respectively).
We show the counts from all 17 SMGs and those obtained when we remove the ten sources with $z_{\rm spec}=3.09$. 
We also summarize the count statistics in Table 4.
For comparison, we also show other ALMA counts in both panels of Fig.~\ref{count} (\citealt{2013MNRAS.432....2K}; \citealt{2015ApJ...807..128S}; \citealt{2016ApJ...822...36O}; \citealt{2014ApJ...795....5O}; \citealt{2016ApJS..222....1F}) by scaling to 1.1~mm assuming a modified black body with typical values for SMGs (spectral index of $\beta$ = 1.5, dust temperature of 35 K (e.g., \citealt{2014MNRAS.438.1267S}) and $z = 2.5$ where necessary).

As discussed in \citet{2015ApJ...815L...8U}, the volume density of the ALMA SMGs at $z=3.09$ in ADF22 is at least about three orders of magnitude greater than the expected value in general fields. 
We now investigate whether the protocluster members affect the number counts or not, considering ALMA SMGs at all redshifts. As shown in Fig.~\ref{count}, we see a possible excess in both of the differential and cumulative counts (for all of 17 SMGs). 
In the case of cumulative counts, the counts in ADF22 are approximately five times higher than those found in ALESS at $S_{\rm 1.1mm}>2$ mJy (\citealt{2013MNRAS.432....2K}) and four to five times higher than those obtained at $S_{\rm 1.1mm} > 1$~mJy (\citealt{2016ApJ...822...36O}; \citealt{2016ApJS..222....1F}). 
The group of SMGs at $z=3.09$ is undoubtedly responsible for this excess.
On the other hand, we do not see an excess in the faintest bin (i.e., sub-mJy sources).
A possible attribution is that the elevation of dusty star-forming activity in dense environment is significant only for relatively bright SMGs ($S_{\rm 1.1mm}\gtrsim1$~mJy). Deeper observation in the future will provide an answer.


Here another question naturally arises. What are the counts in typical fields?
The core of the SSA22 protocluster is an unusual environment.
Therefore we have to identify ALMA SMGs which are not associated with the protocluster and utilize them in calculating the source counts in a typical field.
It is, however, difficult to derive the counts correctly because seven sources with $\ge5\sigma$ do not yet have $z_{\rm spec}$.
Here we conservatively created counts using all seven ALMA SMGs (see Fig.~\ref{count}), some of which may also be at $z=3.09$.
The cumulative source counts from the seven SMGs and previous counts at bright fluxes (\citealt{2013MNRAS.432....2K}; \citealt{2015ApJ...807..128S}) are fitted to a double-power law function of the form, $N(>S)=N'/S' [(S/S')^\alpha+(S/S')^\beta]^{-1}$, which yields the best-fit parameters of $N'=200\pm30$~deg$^{-2}$, $S'=4.9\pm0.2$~mJy, $\alpha=10.4\pm1.6$, and $\beta=1.9\pm0.1$ (the right panel of Fig. \ref{count}).
Our counts excluding known protocluster members are in reasonable agreement with recent estimates which is ought to be relatively free of cosmic variance; those from a number of calibration fields (\citealt{2016ApJ...822...36O}) and those from a 2 arcmin$^{2}$ contiguous survey (\citealt{2016PASJ...68...36H}).
On the one hand, those counts from the seven SMGs seem to be several times lower than \citet{2016ApJS..222....1F} at $S_{\rm 1.1mm}<1$~mJy.
One possible explanation for this discrepancy is that previous counts using 3-4$\sigma$ sources are seriously affected by contaminants, as pointed out by \citet{2016ApJ...822...36O}.
Another explanation involves clustering, although this effect may also contribute to our counts.
The counts derived from serendipitously detected sources around other targeted galaxies might be biased or clustered as mentioned in previous papers (e.g., \citealt{2015ApJ...807..128S}; \citealt{2016ApJS..222....1F}). 

Our results suggest that the cluster environments are recognizable in the SMG number counts. 
At the same time, there are still large uncertainties for submm/mm number counts, including field to field variation, the shape of dust SEDs, and absolute flux uncertainties. Forthcoming much wider ALMA surveys with sufficient depth, covering a variety of environments, will significantly improve our understanding of the submm/mm counts and their dependence on environment.

\section{Summary}

We have presented the results from a 103-pointing ALMA 1.1~mm mosaic in the SSA22 field.
We covered a 7 arcmin$^2$ area contiguously at the junction of a 50-Mpc scale cosmic large-scale structure at $z=3.09$ containing three 1.1~mm sources discovered previously by the AzTEC/ASTE survey.
Observations were conducted in 2014 and 2015 with different array configurations and therefore we created two maps to handle the different angular resolution data; a $1^{\prime\prime}$-- tapered map (FULL/LORES map), and a $0.^{\prime\prime}7$ resolution map (DEEP/HIRES map).
These maps reach a median rms noise of 75~$\mu$Jy beam$^{-1}$  and 60~$\mu$Jy beam$^{-1}$, respectively.
Applying our source detection algorithm to our two maps ,we discover 18 robust ALMA SMGs with SNR$>5$, with 1.1~mm flux density $S_{\rm 1.1mm}=0.44-5.60$~mJy beam$^{-1}$, corresponding to $L_{\rm IR}\simeq8\times10^{11}$ -- $1\times10^{13}L_\odot$ at $z=3$.

Through a comparison between the AzTEC map and the ALMA map of ADF22, we have found that three single-dish (AzTEC) sources are resolved into eight discrete ALMA SMGs. This suggests that multiple dusty galaxies may often contribute to one bright submm/mm source discovered by single-dishes in dense protocluster environments. The flux densities of AzTEC sources are consistent with the sum of the flux densities of ALMA SMGs within errors for two sources, AzTEC14 and AzTEC77, but not for a third (AzTEC1), possibly indicating that there are additional missing faint and/or diffuse dusty galaxies remaining undetected in the ALMA map.
Ten out of 18 SMGs in our sample are known to lie at $z=3.09$. 
The fact that multiple $z=3.09$ ALMA SMGs are found to comprise two of three of the single-dish (AzTEC) sources suggests that at least in dense protocluster environments interactions may be responsible for a significant fraction of observed SMG multiplicity.


The ALMA SMGs are generally resolved in our data. 
We measured the deconvolved source size for nine brightest ALMA SMGs with $>10\sigma$, conducting Gaussian fitting to the ALMA SMGs.
The median size of $0^{\prime\prime}.32^{+0^{\prime\prime}.13}_{-0^{\prime\prime}.06}$ ($2.4^{+1.0}_{-0.4}$ physical kpc at $z=3.09$) agrees well with previous measurements and there is no recognizable difference between the size of $z=3.09$ protocluster members and the ALMA SMGs in other fields.

We derived the 1.1~mm source counts using all 17 SMGs above 5$\sigma$ from the FULL/LORES map and also the seven SMGs without $z_{\rm spec}=3.09$.
The counts of 17 SMGs are about five times higher than the counts in typical fields at $S_{\rm 1.1mm}\gtrsim1$~mJy, which is caused by the $z=3.09$ SMG concentration associated with the protocluster.
On the other hands, we found that the source counts in ADF22 are consistent with recent unbiased ALMA counts when we exclude known members of the $z=3.09$ structure.

In conclusion, we have obtained deep and high angular resolution imaging covering a certain area using ALMA, which demonstrated that there is an unusual number of dusty galaxies residing at the core of the $z=3.09$ protocluster. The observation provides a basis for understanding the formation of dusty galaxies within cosmic large-scale structure in the early universe.

\begin{center}
\begin{deluxetable*}{cccccccc}
\tabletypesize{\scriptsize}
\tablecaption{Summary of Derived Properties for 9 sources with SNR$\ge10$}
\tablewidth{0pt}
\tablehead{
\colhead{ALMA ID} &\colhead{Major axis} & &  \colhead{Minor axis} &  & \colhead{$L_{\rm 8-1000\mu m}$} & \colhead{SFR$_{\rm IR}$} & \colhead{$\Sigma_{\rm SFR}$} \\
 & ($^{\prime\prime}$ )& (kpc) & ($^{\prime\prime}$) & (kpc) & (log ($L_\odot$)) & ($M_\odot$yr$^{-1}$) & ($M_\odot$yr$^{-1}$kpc$^{-2}$ )
}   
\startdata
ADF22.1 & 0.85 $\pm$ 0.04 & 6.5 $\pm$ 0.3 & 0.33 $\pm$ 0.03 & 2.5 $\pm$ 0.2    &   13.0 $_{- 0.1 }^{+ 0.2 }$ &  1100 $_{- 210 }^{+ 830   }$ &  40 \\
ADF22.2* & 0.29 $\pm$ 0.05 & 2.2 $\pm$ 0.4 & 0.05 $\pm$ 0.11 & 0.4 $\pm$ 0.8    &   12.6 $_{- 0.1 }^{+ 0.2 }$ &  400 $_{- 70 }^{+ 300   }$ &  300 \\
ADF22.3 & 0.45 $\pm$ 0.05 & 3.4 $\pm$ 0.4 & 0.18 $\pm$ 0.08 & 1.4 $\pm$ 0.6    &   12.5 $_{- 0.1 }^{+ 0.2 }$ &  370 $_{- 70 }^{+ 280   }$ &  50 \\ 
ADF22.4 & 0.25 $\pm$ 0.05 & 1.9 $\pm$ 0.4 & 0.04 $\pm$ 0.11 & 0.3 $\pm$ 0.8    &   12.6 $_{- 0.1 }^{+ 0.2 }$ &  380 $_{- 70 }^{+ 290   }$ &  420 \\ 
ADF22.5* & 0.57 $\pm$ 0.06 & 4.4 $\pm$ 0.5 & 0.26 $\pm$ 0.06 & 2.0 $\pm$ 0.5    &   12.7 $_{- 0.1 }^{+ 0.2 }$ &  480 $_{- 90 }^{+ 360   }$ &  30 \\
ADF22.6 & 0.38 $\pm$ 0.08 & 2.9 $\pm$ 0.6 & 0.30 $\pm$ 0.14 & 2.3 $\pm$ 1.1    &   12.4 $_{- 0.1 }^{+ 0.2 }$ &  280 $_{- 50 }^{+ 210   }$ &  30 \\
ADF22.7 & 0.32 $\pm$ 0.08 & 2.4 $\pm$ 0.6 & 0.22 $\pm$ 0.10 & 1.7 $\pm$ 0.8    &   12.5 $_{- 0.1 }^{+ 0.2 }$ &  320 $_{- 60 }^{+ 240   }$ &  50 \\
ADF22.8 & 0.26 $\pm$ 0.07 & 2.0 $\pm$ 0.5 & 0.18 $\pm$ 0.05 & 1.4 $\pm$ 0.4    &   12.3 $_{- 0.1 }^{+ 0.2 }$ &  230 $_{- 40 }^{+ 180   }$ &  50 \\
ADF22.9 & 0.28 $\pm$ 0.14 & 2.1 $\pm$ 1.1 & 0.26 $\pm$ 0.17 & 2.0 $\pm$ 1.3    &   12.2 $_{- 0.1 }^{+ 0.2 }$ &  160 $_{- 30 }^{+ 120   }$ &  20 
\enddata
\tablecomments{
Derived properties of the nine sources with SNR $\ge$10.
The second and fourth column show a deconvolved FWHM of the major and minor axis, derived using {\sc imfit}.
The third and fifth column represent corresponding physical scale.
Infrared luminosity ($L_{\rm 8-1000\mu m}$) is estimated using several templates of dusty galaxies as described in \citet{2015ApJ...815L...8U}.
SFR$_{\rm IR}$ are calculated from $L_{\rm 8-1000\mu m}$, using the empirical calibration by \citet{1998ARA&A..36..189K} adjusted for Kroupa initial mass function (\citealt{2001MNRAS.322..231K}).
We also roughly estimated surface density of SFR ($\Sigma_{\rm SFR}$) using SFR$_{\rm IR}$ and derived source size.
* Since ADF22.2 and ADF22.5 don't have $z_{\rm spec}$, we assume $z=3.0$, which is a median redshift of the AzTEC sources (\citealt{2012ApJS..200...10S}).
}
\end{deluxetable*}
\end{center}


\begin{center}
\begin{deluxetable*}{cccccccccc}
\tabletypesize{\scriptsize}
\tablecaption{Differential and cumulative number counts for all SMGs}
\tablewidth{0pt}
\tablehead{
\multicolumn{5}{c}{Differential Counts } &  \multicolumn{5}{c}{Cumulative Counts} \\
\colhead{$S_{\rm 1.1mm}$} &\colhead{N$_{\rm all}$} & \colhead{$dN/dS$} & \colhead{N$_{\rm field}$} & \colhead{$dN/dS$} & \colhead{$S_{\rm 1.1mm}$} &\colhead{N$_{\rm all}$}& \colhead{$N$ ($>S$)} &\colhead{N$_{\rm field}$}& \colhead{$N$ ($>S$)}  \\
(mJy) & & ($10^3$ mJy$^{-1}$ deg$^{-2}$) & & ($10^3$ mJy$^{-1}$ deg$^{-2}$) & (mJy) & &  ($10^3$ deg$^{-2}$) & &  ($10^3$ deg$^{-2}$) 
}   
\startdata
0.64 & 9 & $9.4^{+6.1}_{-2.6}$  & 5 & $5.6^{+4.6}_{-1.9}$   &  0.40 & 17 &  $9.8^{+5.1}_{-2.2}$ & 7 & $4.4^{+3.3}_{-1.3}$\\
1.59 & 7 & $2.4^{+2.1}_{-0.8}$ &  2 & $0.7^{+1.1}_{-0.4}$   & 1.00  & 8 & $4.1^{+3.4}_{-1.4}$ & 2 & $1.0^{+1.7}_{-0.6}$ \\
4.00 & 1 & $0.1^{+0.3}_{-0.1}$ & 0 & ---                               & 2.52 & 1 & $0.5^{+1.2}_{-0.4}$ & 0 & --- 
\enddata
\tablecomments{
We calculated the differential and cumulative counts in two ways: using all 17 SMGs (all) or the seven SMGs without $z_{\rm spec}=3.09$ ({\it field}).
The errors are 1$\sigma$ Poisson errors (\citealt{1986ApJ...303..336G}).
}
\end{deluxetable*}
\end{center}

\acknowledgments

We deeply appreciate the anonymous referee for a significant number of valuable comments.
We are thankful for Ken Mawatari, for helping us to compile redshift catalogs of $z=3.09$ LAEs and LBGs in the literatures.
HU is supported by the ALMA Japan Research Grant of NAOJ Chile Observatory, NAOJ-ALMA-0071, 0131, 140, and 0152. HU is supported by JSPS Grant-in-Aid for Research Activity Start-up (16H06713).
YT is supported by JSPS KAKENHI No. 25102073.
RJI acknowledges support from ERC in the form of the Advanced Investigator Programme, 321302, COSMICISM.
IRS and DMA acknowledge support from STFC (ST/L00075X/1).
IRS acknowledge support from the ERC Advanced Investigator program
DUSTYGAL 321334, and a Royal Society/Wolfson Merit Award. 
SI acknowledge the support of the Netherlands Organization for
Scientific Research (NWO) through the Top Grant Project 614.001.403.
This paper makes use of the following ALMA data: ADS/JAO.ALMA\#2013.1.00162.S. ALMA is a partnership of ESO (representing its member states), NSF (USA) and NINS (Japan), together with NRC (Canada) and NSC and ASIAA (Taiwan) and KASI (Republic of Korea), in cooperation with the Republic of Chile. The Joint ALMA Observatory is operated by ESO, AUI/NRAO and NAOJ.



{\it Facilities:} \facility{ALMA}.

\appendix

\section{A. Tentative Source Catalog}

\begin{figure}
\epsscale{0.6}
\plotone{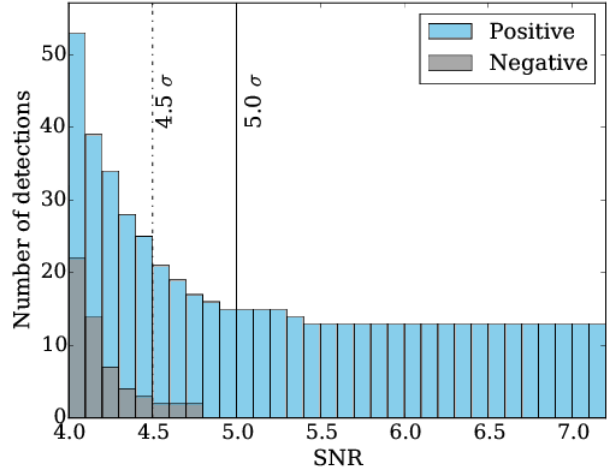}
\caption{
Cumulative number of positive and negative peaks in the DEEP/HIRES map. There are no negative peaks above 5$\sigma$.
We adopt 4.5$\sigma$ as a tentative detection limit for the DEEP/HIRES map.
Above this limit, there is one negative peak.
}
\label{np_d}
\end{figure}

In addition to the robust samples of 18 ALMA SMGs with $>5\sigma$ detection, there are also some possible source candidates with slightly lower significances.
Considering the result from our test using negative maps shown in Fig. \ref{snr} and Fig. \ref{np_d}, we adopt 4.0$\sigma$ and 4.5$\sigma$ as tentative detection thresholds for the FULL/LORES and DEEP/HIRES maps, respectively.
The test suggests that about half of these sources detected tentatively may be false detections and therefore we need to be careful in utilizing the catalog.

\begin{center}
\begin{deluxetable*}{cccccccccccc}
\tabletypesize{\scriptsize}
\tablecaption{Properties of tentative ADF22 sources}
\tablewidth{0pt}
\tablehead{
\colhead{(1)}            & \colhead{(2)}                  &\colhead{(3)}                              & \colhead{(4)}                                & \colhead{(5)}                         & \colhead{(6)} \\
\colhead{ALMA ID}  & \colhead{ALMA NAME} &\colhead{$\sigma_{\rm ALMA}$} & \colhead{S$^{\rm pk}_{\rm ALMA}$}  & \colhead{$S_{\rm ALMA}$}  & \colhead{Map} \\
\colhead{}            & \colhead{}                  &\colhead{($\mu$Jy beam$^{-1}$)}       & \colhead{}                                & \colhead{(mJy) }                         & \colhead{} 
}
\startdata
ADF22. 19 &  ALMAJ221733.87 +001646.1 & 63 &  4.3 &  0.29 $\pm$ 0.06 & FULL/LORES \\
ADF22. 20 & ALMAJ221734.69 +001635.2   & 68  & 4.3&  0.67 $\pm$ 0.09 & FULL/LORES \\
ADF22. 21$^{\dagger}$ &  ALMAJ221733.09 +001718.5  &  78  &  4.3 & 0.79 $\pm$ 0.08  & FULL/LORES \\
ADF22. 22 &  ALMAJ221732.96 +001636.0  &  67  & 4.3 & 0.42 $\pm$ 0.07 & FULL/LORES \\
ADF22. 23 &  ALMAJ221735.47 +001805.1   & 65  & 4.2 & 0.30 $\pm$ 0.06 & FULL/LORES \\
ADF22. 24 &  ALMAJ221737.43 +001723.0  &  67 &  4.1 & 0.56 $\pm$ 0.08 & FULL/LORES \\
ADF22. 25 &  ALMAJ221733.81 +001656.5  &  72  & 4.1 & 0.44 $\pm$ 0.08 & FULL/LORES \\
ADF22. 26 &  ALMAJ221733.19 +001752.7  &  82  & 4.1 & 0.63 $\pm$ 0.07 &FULL/LORES \\
ADF22. 27 &  ALMAJ221737.18 +001832.0  &  85 & 4.0 & 0.25 $\pm$ 0.06 &FULL/LORES \\
ADF22. 28 &  ALMAJ221732.50 +001729.5  &  78  & 4.0 & 0.48 $\pm$ 0.08 & FULL/LORES \\
ADF22. 29 &  ALMAJ221738.35 +001710.8  &  109 & 4.0 & 1.12 $\pm$ 0.11 & FULL/LORES \\
ADF22. 30 &  ALMAJ221732.19 +001642.0  &  63   & 4.7 & 0.29 $\pm$ 0.09 & DEEP/HIRES \\
ADF22. 31 &  ALMAJ221734.97 +001527.6  &  64   & 4.6  & 0.41 $\pm$ 0.12 & DEEP/HIRES \\
ADF22. 32 &  ALMAJ221734.08 +001632.6  &  59   &  4.5 & 0.35 $\pm$ 0.12 & DEEP/HIRES 
\enddata 
\tablecomments{
Columns are generally similar with Table 2. We select 11 sources with $SNR\ge4.0$ using the FULL/LORES map.
We also include three sources with $SNR\ge4.5$ (but not selected using the FULL/LORES map) on the basis of the DEEP/HIRES map.
In total 14 tentative sources are found. $^{\dagger}$: CO(4-3) line at $z=0.71$ is detected and identified using the band 6 cube (Hayatsu et al., in preparation).
}
\end{deluxetable*}
\end{center}

\section{B. Catalogs of proto-cluster members in ADF22}

\begin{center}
\begin{deluxetable*}{cccccccccccc}
\tabletypesize{\scriptsize}
\tablecaption{Known member galaxies of the $z=3.09$ proto-cluster in ADF22}
\tablewidth{0pt}
\tablehead{
\colhead{(1)}         & \colhead{(2)}                           &\colhead{(3)}                     & \colhead{(4)}     & \colhead{(5)} & \colhead{(6)} \\
\colhead{Galaxy}  & \colhead{Coordinate (J2000)} &\colhead{$z_{\rm spec}$} & \colhead{Type}  & \colhead{IDs} & \colhead{ALMA source}
}
\startdata
LAEs \\
001 & 22:17:33.10  +00:18:29.0 & 3.090 & Ly$\alpha$ & (M05) & --- \\
002 & 22:17:35.61  +00:18:00.2 & 3.091 & Ly$\alpha$ & 016 (N13) & ---  \\
003 & 22:17:31.73  +00:16:06.9 & 3.101 & Ly$\alpha$ & 023 (N13) & ---  \\
004 & 22:17:34.17  +00:16:09.7 & 3.096 & Ly$\alpha$ & 024 (N13) & ---  \\
005 & 22:17:36.74  +00:16:28.8 & 3.091 & Ly$\alpha$ & 025 (N13)  & --- \\
006 & 22:17:31.80  +00:17:17.9 & 3.088 & Ly$\alpha$ & 028 (N13) & ---  \\
007 & 22:17:33.63  +00:17:15.1 & 3.092 & Ly$\alpha$ & 031 (N13)  & --- \\
008 & 22:17:34.77  +00:15:41.3 & 3.099 & Ly$\alpha$ & 038 (N13) & ---  \\
009 & 22:17:35.97  +00:16:30.2 & 3.094 & Ly$\alpha$ & 045 (N13)  & --- \\
010 & 22:17:34.70  +00:16:33.4 & 3.090 & Ly$\alpha$ & 053 (N13) & ---  \\
011 & 22:17:34.10  +00:15:40.2 & 3.101 & Ly$\alpha$ & 061 (N13) & ---  \\
012 & 22:17:31.24  +00:17:32.1 & 3.084 & Ly$\alpha$ & 072 (N13)  & --- \\
       &                                          & 3.0845 & \textsc{[Oiii]}$\lambda$5007 & 072 (E14)  & --- \\
013 & 22:17:37.68  +00:16:48.3 & 3.090 & Ly$\alpha$ & 078 (N13) & ---  \\
       &                                          & 3.0870 & \textsc{[Oiii]}$\lambda$5007 & 078 (E14) & ---  \\
014 & 22:17:35.44  +00:16:47.6 & 3.087 & Ly$\alpha$ & 082 (N13)  & --- \\
       &                                          & 3.0873 & \textsc{[Oiii]}$\lambda$5007 & 082 (E14) & ---  \\
015 & 22:17:36.14  +00:15:40.7 & 3.095 & Ly$\alpha$ & 091 (N13) & ---  \\
016 & 22:17:31.14  +00:16:42.9 & 3.096 & Ly$\alpha$ & 111 (N13)  & --- \\
017 & 22:17:32.72  +00:15:54.2 & 3.096 & Ly$\alpha$ & 112 (N13) & ---  \\
018 & 22:17:33.46  +00:17:01.2 & 3.093 & Ly$\alpha$ & 115 (N13)  & --- \\
019 & 22:17:32.84  +00:16:48.8 & 3.092 & Ly$\alpha$ & 130 (N13)  & --- \\
\\
LBGs \\
001 & 22:17:31.49 +00:16:31.2 & 3.098 & Ly$\alpha$ & M25 (N13)  & --- \\
002 & 22:17:31.66 +00:16:58.0 & 3.094 & Ly$\alpha$ & M28, 012 (S03; N13)  & --- \\
       &                                         & 3.0902 & \textsc{[Oiii]}$\lambda$5007 & 012 (E14)  & --- \\
003 & 22:17:36.87 +00:17:12.4 & 3.099 & Ly$\alpha$ & M31 (N13)  & --- \\
004 & 22:17:33.80 +00:17:57.2 & 3.084 & Ly$\alpha$ & M34 (N13)  & --- \\
005 & 22:17:37.66 +00:18:20.9 & 3.086 & Ly$\alpha$ (abs) & C50 (S03)  & --- \\
\\
$K$-band selected galaxies \\
001 & 22:17:37.1 +00:17:12.4 & 3.0899 &\textsc{\textsc{[Oiii]}}~$\lambda$5007 & (K15)  & ADF22.8 \\
002 & 22:17:37.3 +00:16:30.7& 3.0888 & \textsc{\textsc{[Oiii]}}~$\lambda$5007 & (K15) & ---  \\
003 & 22:17:36.5 +00:16:22.6 & 3.0945 & \textsc{\textsc{[Oiii]}}~$\lambda$5007 & (K15) & ADF22.9  \\
004 & 22:17:31.8 +00:16:06.3 & 3.0981 & \textsc{\textsc{[Oiii]}}~$\lambda$5007 & (K15) & ---  \\
005 & 22:17:32.0 +00:16:55.5 & 3.0909 &\textsc{\textsc{[Oiii]}}~$\lambda$5007 & (K15) & ADF22.12  \\
006 & 22:17:37.3 +00:18:23.2 & 3.0851 & \textsc{\textsc{[Oiii]}}~$\lambda$5007 & K15a (K15; K16) & ---  \\
007 & 22:17:36.8 +00:18:18.2 & 3.0854 &\textsc{\textsc{[Oiii]}}~$\lambda$5007 & K15b (K15; K16) & ADF22.16  \\
008 & 22:17:37.1 +00:18:17.9 & 3.0774 & \textsc{\textsc{[Oiii]}}~$\lambda$5007 & K15d (K15; K16) & ---  \\
009 & 22:17:37.1 +00:18:22.4 & 3.0925 &\textsc{\textsc{[Oiii]}}~$\lambda$5007 & K15e (K15; K16) & ADF22.11 \\
010 & 22:17:36.9 +00:18:38.0 & 3.0866& \textsc{\textsc{[Oiii]}}~$\lambda$5007 & K15f (K15; K16)  & --- \\
\\
X-ray sources \\
and LABs \\
001 & 22:17:32.0 +00:16:55.6 & 3.091 & \textsc{\textsc{[Oiii]}}~$\lambda$5007 & 114 (L09), LAB12 (M04; G09) & ADF22.12 \\ 
002 & 22:17:32.2,+00:17:36.0 & 3.097 & \textsc{[Cii]}~158$\mu$m & 116 (L09) & ADF22.7 \\ 
003 & 22:17:32.4,+00:17:43.9 & 3.092 & CO(3-2) & 120 (L09)  & ADF22.1 \\ 
004 & 22:17:35.8,+00:15:59.1 & 3.089 & Ly$\alpha$ & 139 (L09), LAB14 (M04; G09) & ADF22.6 \\ 
005 & 22:17:36.5,+00:16:22.6 & 3.084 & Ly$\alpha$ & 140 (L09), QSO (S98) & ADF22.9 \\ 
006 & 22:17:37.0,+00:18:20.8 & 3.091 & CO(9-8), \textsc{[Cii]}~158$\mu$m &142 (L09) & ADF22.4 \\ 
007 & 22:17:37.3,+00:16:30.7 & 3.0888 & \textsc{\textsc{[Oiii]}}~$\lambda$5007 & 144 (L09) & --- \\ 
008 & 22:17:37.3,+00:18:23.5 & 3.0851 & \textsc{\textsc{[Oiii]}}~$\lambda$5007 &145 (L09) & ---  
\enddata 
\tablecomments{
(1) Galaxy population and IDs in this work. There are overlaps between K-band selected galaxies and X-ray sources/LABs.
(2) Coordinates in the literatures.
(3), (4) Spectroscopic redshifts and the line used to determine redshifts.
(5) IDs in the literatures as well as the literatures (\citet{2005ApJ...634L.125M} (M05); \citet{2013ApJ...765...47N} (N13); \citet{2014ApJ...795...33E} (E14); \citet{2003ApJ...592..728S} (S03);  \citet{2015ApJ...799...38K} (K15); \citet{2016MNRAS.455.3333K} (K16); \citet{2009MNRAS.400..299L} (L09); \citet{2004AJ....128..569M} (M04); \citet{2009ApJ...700....1G} (G09)). The redshift determination of X-ray sources and ALMA sources are described in \citet{2015ApJ...815L...8U} and this work.
(6) IDs of ALMA counterparts in ADF22.
}
\end{deluxetable*}
\end{center}

\bibliographystyle{apj}





\end{document}